\documentclass[
    english,10pt,aps,prb,twocolumn,nofootinbib,amsmath,amssymb,groupedaddress]{revtex4-2}

\usepackage{silence}        
\usepackage[T1]{fontenc}    
\usepackage{lmodern}        
\usepackage{babel}          
\usepackage[
    babel
    ]{microtype}            
\usepackage[all]{nowidow}   

\usepackage{mathtools}  
\usepackage{amstext}    
\usepackage{amssymb}    
\usepackage{bm}         
\usepackage{bbold}      
\usepackage{aligned-overset}    

\usepackage{physics}    

\usepackage{csquotes}   
\usepackage{xcolor}     
\usepackage{lipsum}     
\usepackage[            
    normalem            
    ]{ulem}
\usepackage{graphicx}   
\usepackage[
    inkscapearea=page
    ]{svg}              
\usepackage[
    off,                
    clean,              
    extractpath=svgdir
    ]{svg-extract}      
\usepackage{booktabs}   

\usepackage[
    bookmarks=true,
    colorlinks,
    citecolor=blue,
    linkcolor=blue,
    urlcolor=blue
    ]{hyperref}                     
\usepackage[capitalize]{cleveref}   

\definecolor{darkred}{RGB}{165,50,50}
\definecolor{darkgreen}{RGB}{0,100,0}
\definecolor{darkblue}{RGB}{50,50,165}
\definecolor{lightgray}{RGB}{211,211,211}

\DeclareMathOperator{\ee}{\mathrm e}
\DeclareMathOperator{\TT}{\mathrm T}
\DeclareMathOperator{\dc}{\mathrm dc}
\DeclareMathOperator{\conn}{\mathrm conn}

\newcommand{\qftop}[1]{\hat{#1}}
\newcommand{\nn}{\qftop{n}}
\newcommand{\cc}{\qftop{c}^{\vphantom{\dagger}}}
\newcommand{\cdag}{\qftop{c}^\dagger}
\newcommand{\ff}{\qftop{f}^{\vphantom{\dagger}}}
\newcommand{\fdag}{\qftop{f}^\dagger}
\newcommand{\cg}{c^{\vphantom{+}}}  
\newcommand{\cgdag}{c^{+}}          
\renewcommand{\AA}{\qftop{A}}       
\newcommand{\xx}{\qftop{\sigma}_x}
\newcommand{\yy}{\qftop{\sigma}_y}
\newcommand{\zz}{\qftop{\sigma}_z}

\newcommand{\calZ}{\mathcal{Z}}
\newcommand{\calD}{\mathcal{D}}
\newcommand{\calS}{\mathcal{S}}

\newcommand{\ii}{\mathrm{i}}
\newcommand{\HH}{\qftop{H}}
\newcommand{\HHH}{\qftop{\mathcal{H}}}

\newcommand{\vect}[1]{\bm{\mathrm{#1}}}
\newcommand{\matr}[1]{\underline{\underline{#1}}}
\newcommand{\one}{\mathbb{1}}
\newcommand{\up}{\uparrow}
\newcommand{\down}{\downarrow}
\newcommand{\wbar}[1]{\overline{#1}}
\newcommand{\pbs}{\;}                   
\newcommand{\nbs}{\!\!\!\:}             

\newcommand{\chivertex}{\chi_{\text{vertex}}}
\newcommand{\chione}[1][]{\chi^{(1) \, #1}}
\newcommand{\chitwo}[1][]{\chi^{(2) \, #1}}
\newcommand{\chicon}{\chi}
\newcommand{\chinnn}{\chi_{nnn}}
\newcommand{\chinxx}{\chi_{nxx}}
\newcommand{\chinyy}{\chi_{nyy}}
\newcommand{\chinzz}{\chi_{nzz}}
\newcommand{\chixyz}{\chi_{xyz}}

\newcommand{\chitwovertex}[1][]{\chitwo[#1]_{\text{vertex}}}
\newcommand{\corr}{X}
\newcommand{\Pthree}{P}
\newcommand{\uev}[1]{\ev*{\underline{\TT #1}}}  

\renewcommand{\t}[1]{\text{#1}}

\makeatletter
\newcommand{\thefontsize}{\f@size pt\par}
\makeatother

\newcommand{\program}[1]{\texttt{#1}}

\graphicspath{{./Plots/}{./Diagrams/}}
\svgpath{{./Diagrams/}}

\newcommand{\includeinkscapefigure}[2][]{\includegraphics[#1]{#2}}

\allowdisplaybreaks[1]    
\newcommand{\Title}{Non-linear responses and three-particle correlators in correlated electron systems exemplified by the Anderson impurity model}
\newcommand{\Subtitle}{Chi2 Paper}
\newcommand{\Author}{Patrick Kappl, Friedrich Krien, Clemens Watzenb\"ock, Karsten Held}
\newcommand{\Keywords}{chi2, three-particle, 3-particle, Raman, nonlinear response, DMFT, AIM, Hubbard model }

\hypersetup{
    pdfauthor = {\Author},
    pdftitle = {\Title},
    pdfsubject = {\Subtitle},
    pdfkeywords = {\Keywords},
}

\begin{document}
    \title{\Title}
    \author{\Author}
    \affiliation{Institute of Solid State Physics, TU Wien, 1040 Vienna, Austria}
    \date{\today}

    \begin{abstract}
    Three-particle correlators are relevant for, among others, Raman, Hall and non-linear
    responses. They are also required for the next order of approximations extending
    dynamical mean-field theory diagrammatically. We present a general formalism on how to
    treat these three-particle correlators and susceptibilities, and calculate the local
    three-particle response of the Anderson impurity model numerically. We find that
    genuine three-particle vertex corrections are sizable. In particular, it is not
    sufficient to just take the bare bubble terms or corrections based on the two-particle
    vertex. The full three-particle vertex must be considered.
\end{abstract}
 
    \maketitle

    \section{Introduction}
\label{sec:introduction}

Our physical understanding is very much based on one-particle and two-particle Green's
functions, upon which books on quantum field theory generally focus~\cite{Abrikosov1975}.
On the one-particle level, we understand quasiparticle renormalizations and life times,
metal--insulator transitions as well as magnetic ordering in the symmetry-broken phase.
The one-particle Green's function and self-energy are also at the heart of dynamical mean
field theory (DMFT) \cite{Metzner1989,Jarrell1992,Georges1992,Georges1996}, which
calculates the self-energy by (self-consistently) summing up the local contribution of all
Feynman diagrams~\cite{Metzner1989}. Hence, it is maybe not surprising that the success of
DMFT had a focus on describing the aforementioned one-particle properties such as
quasiparticle renormalizations and the Mott--Hubbard metal--insulator transition.

On the two-particle level, we have the two-particle Green's function from which we can
calculate physical responses such as the magnetic or charge susceptibility. Here, the
two-particle vertex plays the role of the self-energy. It describes all physics beyond a
rather trivial bare bubble susceptibility which is akin to the non-interacting case, only
now with renormalized one-particle Green's function lines. On this two-particle level
there has been some recent progress to describe electronic correlations -- brought about
through diagrammatic extensions of DMFT
\cite{Kusunose2006,Toschi2007,Rubtsov2008,Rohringer2013,Taranto2014,Ayral2015,Li2015}; for
a review see Ref.~\cite{RMPVertex}. These start from a local vertex that encodes all DMFT
correlations and subsequently generate non-local correlations through the Bethe--Salpeter
equation or parquet equations. Quite naturally these extensions allow for a better
description of two-particle quantities such as the (quantum) critical behavior in the
vicinity of a phase transition
\cite{Rohringer2011,Antipov2014,Hirschmeier2015,Schaefer2016}, spin-fluctuation induced
pseudogaps \cite{Katanin2009,Schaefer2015-2,Krien2021} and superconducting instabilities
\cite{Otsuki2014,Kitatani2015}.

The next level, the three-particle Green's function and vertices, are hitherto
by-and-large a blank spot in our understanding of strongly correlated electron systems.
First results for the diagrammatic extensions of DMFT~\cite{Ribic2017b} show that
three-particle vertices are, at least in some parameter regimes of the Hubbard model,
relevant. While our physical understanding and intuition is presently much more based on
the one- and two-particle physics, there are also physical processes that are generically
connected to three-particle correlators:

Take for example Raman scattering, with an incoming and outgoing light frequency and a
transferred phonon frequency. These three bosonic frequencies are connected to three
electrons (particles), each with one creation and one annihilation operator. The same
applies to the Hall response, i.e., the off-diagonal conductivity in a magnetic field. The
conductivity by itself is a two-particle correlator in the Kubo formalism of linear
response~\cite{Kubo1957}. Considering small magnetic fields, these can be treated in
linear response as well, making the Hall coefficient a three-particle correlator
altogether. In principle, calculating these observables requires the calculation of the
full three-particle correlator. But hitherto either only a bare bubble-like diagram is
taken or corrections based on the two-particle vertex are included, see
e.g.~\cite{Jorio2011,Rostami2017,Vandelli2019}.

Another research area which is the domain of correlators with more than two particles is
non-linear response \cite{Kubo1957,Rostami2021NLR}. These responses are in general weaker
than linear responses and often the most relevant correction is thus the second-order
response that is connected to a three-particle correlator. Interestingly,
Refs.~\cite{PhysRevB.103.195133,PhysRevB.104.085151} found that correlation effects can
enhance non-linear responses in strongly correlated electron systems, but, again, they
only took into account one-particle renormalizations; full vertex corrections were still
neglected.

Three-particle correlators are also employed for calculating two-particle correlators
reliably using the so-called improved estimators based on the equation of
motion~\cite{Hafermann2012,Hafermann2014,Gunacker2016,Moutenet2018,Kaufmann2019}.

Against this background, it is the aim of the present paper to do some first steps in
computing, analyzing and understanding these three-particle correlators. Specifically, we
consider correlators of three bosonic operators and thus three time arguments (or two time
differences or frequencies). For the sake of simplicity and for keeping the numerical
effort manageable, we concentrate on local correlators of an Anderson impurity model
(including one at DMFT self-consistency for the two-dimensional Hubbard model). Since the
Raman and Hall response couple to light through non-local fermionic operators, we here
focus on the non-linear response described by local operators only. There are just three
local non-vanishing three-particle correlators (and symmetrically related ones): a
second-order density susceptibility ($nnn$) with three density operators, a mixed
density-magnetic susceptibility ($nzz$) describing the second-order response of the
density to a magnetic field in $z$ direction, and a chiral susceptibility ($xyz$). The
latter corresponds to a correlator with one spin in all three directions. These are
arranged like thumb, index and middle finger of the right hand and hence chiral according
to the definition introduced by Kelvin in 1884~\cite{Kelvin1904} (not invariant under any
mirror transformation). Such a chiral susceptibility arises in the continuity equation of
the $t$-$J$ model or in presence of the direct exchange interaction~\cite{Krien2017}.

The outline of the paper is as follows: In \cref{sec:theory} we give a very brief
introduction to response theory and define all necessary two- and three-particle
quantities as well as the relationships between them and the response functions.
\Cref{sec:models} describes the models we use in our calculations and why we chose them.
The numerical results are then presented and analyzed in \cref{sec:results}. Finally, in
\cref{sec:conclusion,sec:outlook} we give a conclusion and outlook.
     \section{Theory}
\label{sec:theory}

In general, response theory describes the relation between cause and effect. For our
purposes this boils down to quantifying how the expectation value of an arbitrary, bosonic
operator $\ev*{\AA_i}$ depends on some external \enquote{force} $F_j$. As shown in detail
in \cref{sec:nlrt} this can be studied by expanding $\ev*{\AA_i}$ in a functional Taylor
series:
\begin{equation}
    \begin{split}
        \ev*{\AA_i(\tau)}_{\vect{F}} = {} & \ev*{\AA_i(\tau)}_{\vect{F}=0} \\
            & {} + \sum_j \int_0^\beta \dd{\tau'} F^{\vphantom{(}}_j(\tau')
                \chi^{(1)}_{ji}(\tau', \tau) \\
            & {} + \frac{1}{2} \sum_{j k} \int_0^\beta \int_0^\beta \dd{\tau'} \dd{\tau''}
                F_j(\tau') F_k(\tau'') \\
            & {} \quad \times \chitwo_{jki}(\tau', \tau'', \tau) + \dots
    \end{split}
    \label{eq:functional-taylor-series}
\end{equation}
Here, we express everything in imaginary time $\tau$ which runs from zero to $\beta=1/T$,
the inverse temperature. We call the expansion coefficients $\chione$ and $\chitwo$ the
first-order or linear, and second-order or non-linear response function, respectively.
They are simply functional derivatives of $\ev*{\AA_i}$ with respect to $F_j$ and
according to
\cref{eqn:Appendix:NLRTI:ConnectedSusceptibilityOrder1,eqn:Appendix:NLRTI:ConnectedSusceptibilityOrder2}
read
\begin{equation}
    \begin{split}
        \chione_{ij}(\tau, \tau') = {}
            & \eval{\fdv{F_i(\tau)} \ev*{\AA_j(\tau')}}_{\vect{F}=0} \\
        = {}
            & \ev*{\TT \AA_i(\tau) \AA_j(\tau')} - \ev*{\AA_i} \ev*{\AA_j},
    \end{split}
    \label{eq:chi}
\end{equation}
for the first order, and
\begin{multline}
    \chitwo_{ijk}(\tau, \tau', \tau'') = {}
        \eval{\fdv{F_i(\tau)} \fdv{F_j(\tau')} \ev*{\AA_k(\tau'')}}_{\vect{F}=0} \\
    \begin{aligned}
        = {}
            & \ev*{\TT \AA_i(\tau) \AA_j(\tau') \AA_k(\tau'')} \\
            & {} - \ev*{\AA_i} \chi_{jk}(\tau', \tau'')
                - \ev*{\AA_j} \chi_{ik}(\tau, \tau'') \\
            & - \ev*{\AA_k} \chi_{ij}(\tau, \tau') - \ev*{\AA_i} \ev*{\AA_j} \ev*{\AA_k}.
    \end{aligned}
    \label{eq:chi-2}
\end{multline}
for the second order. Here, $\TT$ is the time ordering operator and $\AA_i$ are the
bosonic operators that the external fields $F_i$ couple to, i.e., the Hamiltonian contains
a perturbation term of the form $-\sum_i \AA_i F_i$. If not indicated otherwise the
expectation values are computed with respect to the unperturbed Hamiltonian. From now on
we also drop the superscript denoting the order of the response function whenever the
(number of) arguments allow to infer it.

In Matsubara space the linear response function reads
\begin{equation}
    \begin{split}
        \chi_{ij}^\omega = {}
            & \int_0^\beta \chi^{\vphantom{\omega}}_{ij}(\tau)
            \ee^{\ii \omega \tau} \dd{\tau} \\
        = {}
            & \ev*{\TT \AA_i(\tau) \AA_j}^\omega -
        \delta_{\omega 0} \beta \ev*{\AA_i} \ev*{\AA_j},
    \end{split}
    \label{eq:chi-omega}
\end{equation}
while the non-linear response function is given by
\begin{equation}
    \begin{split}
        \chi_{ijk}^{\omega_1 \omega_2} = {}
            & \int_0^\beta \int_0^\beta \chi^{\vphantom{\omega_1}}_{ijk}(\tau_1, \tau_2)
            \ee^{\ii (\omega_1 \tau_1 + \omega_2 \tau_2)} \dd{\tau_1} \dd{\tau_2} \\
        = {}
            & \ev*{\TT \AA_i(\tau_1) \AA_j(\tau_2) \AA_k}^{\omega_1 \omega_2}
                - \delta_{\omega_1 0} \beta \ev*{\AA_i} \chi_{jk}^{\omega_2} \\
            & {} - \delta_{\omega_2 0} \beta \ev*{\AA_j} \chi_{ik}^{\omega_3}
                - \delta_{\omega_3 0} \beta \ev*{\AA_k} \chi_{ij}^{\omega_1} \\
            & {} - \delta_{\omega_1 0} \delta_{\omega_2 0} \beta^2
                \ev*{\AA_i} \ev*{\AA_j} \ev*{\AA_k}.
    \end{split}
    \label{eq:chi-2-omega}
\end{equation}
Here, we use time-translation invariance to effectively get rid of one imaginary time
argument, $\omega_3 = -\omega_1 - \omega_2$, and the frequency superscript for the
expectation values indicates the Fourier transform of the corresponding imaginary time
expressions defined in \cref{eq:chi,eq:chi-2}.

We see that the response functions are nothing but two- and three-particle correlators
minus their disconnected terms.

So far everything is formulated with general, bosonic operators $\AA_i$. For the rest of
this paper we are, however, only interested in the cases where those are density and spin
operators:
\begin{align}
    \nn & = \nn_\up + \nn_\down = \cdag_\up \cc_\up + \cdag_\down \cc_\down
    \label{eq:n}
    \\
    \qftop{\sigma}_i
        & = \mqty(\cdag_\up & \cdag_\down) \matr{\sigma}_i \mqty(\cc_\up \\ \cc_\down)
    \label{eq:sigma}
\end{align}
with fermionic creation and annihilation operators $\cdag$ and $\cc$ as well as Pauli
matrices $\matr{\sigma}_i$. In the Hamiltonian they couple as $\epsilon \nn$, and $-
\vect{h} \vect{\qftop{\sigma}}$ to (the change of) the one-particle energy $\epsilon$ and
magnetic field $\vect{h}$.

Let us further introduce the following compact notation for the full, bosonic,
two-particle, density and spin correlators
\begin{align}
    \corr_{\sigma_1 \dots \sigma_4}(\tau) & =
        \ev*{\TT \cdag_{\sigma_1}(\tau^+) \cc_{\sigma_2}(\tau)
             \cdag_{\sigma_3}(0^+) \cc_{\sigma_4}(0)},
    \label{eq:2p-spin-corr}
    \\
    \corr_{\alpha \beta} & =
        \!\!\! \sum_{\sigma_1 \dots \sigma_4} \!\!
        s_\alpha^{\sigma_1 \sigma_2} \, s_\beta^{\sigma_3 \sigma_4} \,
        \corr_{\sigma_1 \dots \sigma_4},
    \label{eq:2p-corr}
    \\
    \matr{s}_\alpha & =
    \begin{cases}
        \one, & \alpha = n, \\
        \matr{\sigma}_\alpha, & \alpha \in \{x, y, z\},
    \end{cases}
    \label{eq:s}
\end{align}
where $\tau^+ = \lim_{\varepsilon \rightarrow 0} \tau + \varepsilon$ and $0^+ =
\lim_{\varepsilon \rightarrow 0} 0 + \varepsilon$. Analogously, on the three-particle
level we define
\begin{align}
    \begin{split}
        \corr_{\sigma_1 \dots \sigma_6}(\tau_1^{\vphantom +}, \tau_2^{\vphantom +}) & \\
        \MoveEqLeft[6]
        = \!\! \ev*{\TT \cdag_{\sigma_1}(\tau_1^+) \cc_{\sigma_2}(\tau_1^{\vphantom +})
            \cdag_{\sigma_3}(\tau_2^+) \cc_{\sigma_4}(\tau_2^{\vphantom +})
            \cdag_{\sigma_5}(0^+) \cc_{\sigma_6}(0)} \! ,
    \end{split}
    \label{eq:3p-spin-corr}
    \\
    \corr_{\alpha \beta \gamma} & =
        \!\!\! \sum_{\sigma_1 \dots \sigma_6} \!\!
        s_\alpha^{\sigma_1 \sigma_2} \, s_\beta^{\sigma_3 \sigma_4} \,
        s_\gamma^{\sigma_5 \sigma_6} \, \corr_{\sigma_1 \dots \sigma_6}.
    \label{eq:3p-corr}
\end{align}
In this notation the response functions read
\begin{align}
    \chi_{\alpha \beta} = \conn \corr_{\alpha \beta},
    &&
    \chi_{\alpha \beta \gamma} = \conn \corr_{\alpha \beta \gamma},
    \label{eq:chi-conn-corr}
\end{align}
where $\conn$ denotes only fully connected terms.

On the two-particle level, only the response of the density to changes of the one-particle
energy, $\chi_d$, and the response of the magnetization to changes of the magnetic field
in the same direction, $\chi_m$, do not vanish:\footnote{The extra minus in
\cref{eq:chi-d} is necessary because the density couples with $+\epsilon \nn$ a derivative
with respect to $\epsilon$ brings down $-\nn$ [see \cref{sec:nlrt}], but according to
\cref{eq:chi-conn-corr} we want $\chi_d = \chi_{nn} = +\conn \ev*{\TT \nn \nn}$.}
\begin{align}
    \chi_d(\tau) & = -\fdv{\ev*{\nn}}{\epsilon(\tau)} = \chi_{nn}(\tau),
    \label{eq:chi-d}
    \\
    \chi_m(\tau) & = \fdv{\ev*{\qftop{\sigma}_i}}{h_i(\tau)} = \chi_{ii}(\tau),
        \ \text{with} \ i = x,y,z.
    \label{eq:chi-m}
\end{align}
Using \cref{eq:chi,eq:2p-spin-corr,eq:2p-corr} we can derive well-known relations for the
linear response functions:
\begin{align}
    \begin{split}
        \chi_{nn}(\tau)
        & = \corr_{nn}(\tau) - \! \ev*{\nn}^2 \\
        & = 2 (\chi_{\up \up}(\tau) + \chi_{\up \down}(\tau)),
    \end{split}
    \label{eq:chi-nn}
    \\
    \begin{split}
        \chi_{zz}(\tau)
        & = \corr_{zz}(\tau) = 2 (\chi_{\up \up}(\tau) - \chi_{\up \down}(\tau)),
    \end{split}
    \label{eq:chi-zz}
\end{align}
where the spin susceptibilities are $\chi_{\sigma \sigma'} =
-\fdv*{\ev{\nn_{\sigma'}}}{\epsilon_{\sigma}}$ with $\epsilon_{\sigma}$ denoting the
(change of) the one-particle energy for spin $\sigma$ only, and we assume SU(2) symmetry.

On the three-particle level only the following response functions do not
vanish:\footnote{The extra minus in \cref{eq:chi-nzz} is there for the same reason as in
\cref{eq:chi-d}}
\begin{align}
    \chinnn(\tau_1, \tau_2)
    & = \fdv{\epsilon(\tau_1)} \fdv{\ev{\nn}}{\epsilon(\tau_2)}
        = \fdv{\chi_d(\tau_2)}{\epsilon(\tau_1)},
    \label{eq:chi-nnn}
    \\
    \begin{split}
        \chinzz(\tau_1, \tau_2)
        & = \chinxx(\tau_1, \tau_2) = \chinyy(\tau_1, \tau_2) \\
        & = -\fdv{\epsilon(\tau_1)}\fdv{\ev{\qftop{\sigma}_i}}{h_i(\tau_2)}
            = -\fdv{\chi_m(\tau_2)}{\epsilon(\tau_1)},
    \end{split}
    \label{eq:chi-nzz}
    \\
    \chixyz(\tau_1, \tau_2)
    & = \fdv{h_x(\tau_1)}\fdv{\ev{\zz}}{h_y(\tau_2)}.
    \label{eq:chi-xyz}
\end{align}
We call them the second-order density, density-magnetic, and chiral response functions,
respectively. In \cref{sec:3p-details} we derive relations similar to
\cref{eq:chi-nn,eq:chi-zz} for them. They show, e.g., that $\corr_{xyz}$ contains no
disconnected terms that need to be subtracted -- just like $\corr_{zz}$ on the
two-particle level. Furthermore, in the special case of half-filling, i.e., $n_\sigma =
1/2 = 1 - n_\sigma$ we show that $\chinnn$ and $\chinzz$ vanish.

Remembering the usual definition of an $n$-particle Green's function
\begin{multline}
    G^n_{\sigma_1 \sigma_2 \dots \sigma_{2 n}}(\tau_1, \tau_2, \dots, \tau_{2 n}) \\
    = (-1)^n \ev*{\TT \cc_{\sigma_1}(\tau_1) \cdag_{\sigma_2}(\tau_2) \dots
                  \cdag_{\sigma_{2 n}}(\tau_{2 n})},
    \label{eq:g-n}
\end{multline}
and looking at
\cref{eq:2p-spin-corr,eq:2p-corr,eq:3p-spin-corr,eq:3p-corr,eq:s,eq:chi-conn-corr} we see
that the response functions are basically given by the connected parts of equal time
Green's functions. We only have to take care of the different order of creation and
annihilation operators
\begin{align}
    \corr_{\sigma_1 \dots \sigma_4}(\tau) & =
        G^2_{\sigma_2 \sigma_1 \sigma_4 \sigma_3}(\tau, \tau^+, 0, 0^+),
    \label{eq:corr-g2}
    \\
    \begin{split}
        \corr_{\sigma_1 \dots \sigma_6}(\tau_1, \tau_2) & \\
        \MoveEqLeft[3]
        {} = G^3_{\sigma_2 \sigma_1 \sigma_6 \sigma_5 \sigma_4 \sigma_3}(
            \tau_1^{\vphantom +}, \tau_1^+, 0, 0^+, \tau_2^{\vphantom +}, \tau_2^+).
    \end{split}
    \label{eq:corr-g3}
\end{align}
Let us finally also define the Fourier transform,
\begin{equation}
    \corr_{\sigma_1 \dots \sigma_6}^{\omega_1 \omega_2} =
        \int_0^\beta \! \int_0^\beta \! \corr_{\sigma_1 \dots \sigma_6}(\tau_1,\tau_2)
        \ee^{\ii \omega_1 \tau_1 + \ii \omega_2 \tau_2} \dd{\tau_1} \dd{\tau_2},
    \label{eq:chiomom}
\end{equation}
where $\omega_1$ and $\omega_2$ are bosonic Matsubara frequencies.

\begin{figure}
    \centering
    \includeinkscapefigure{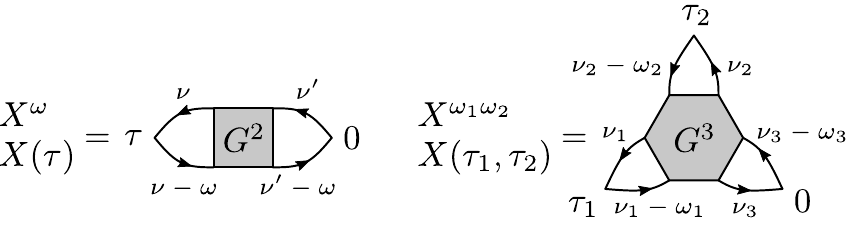}
    \caption{Diagrammatic representation of the full two- and three-particle correlators.
    For brevity the time and frequency labels are given in the same diagrams.}
    \label{fig:correlators}
\end{figure}

With this we can give a diagrammatic representation of the correlators and response
functions. \Cref{fig:correlators} shows the diagrams for the full two- and three-particle
correlators with time and frequency labels.
A particle--hole notation is chosen for the latter. (See
\cref{sec:frequency-notations-of-3p-diagrams} for a detailed look at all 15 frequency
notations of the three-particle Green's function.)

\begin{figure*}
    \centering
    \includeinkscapefigure{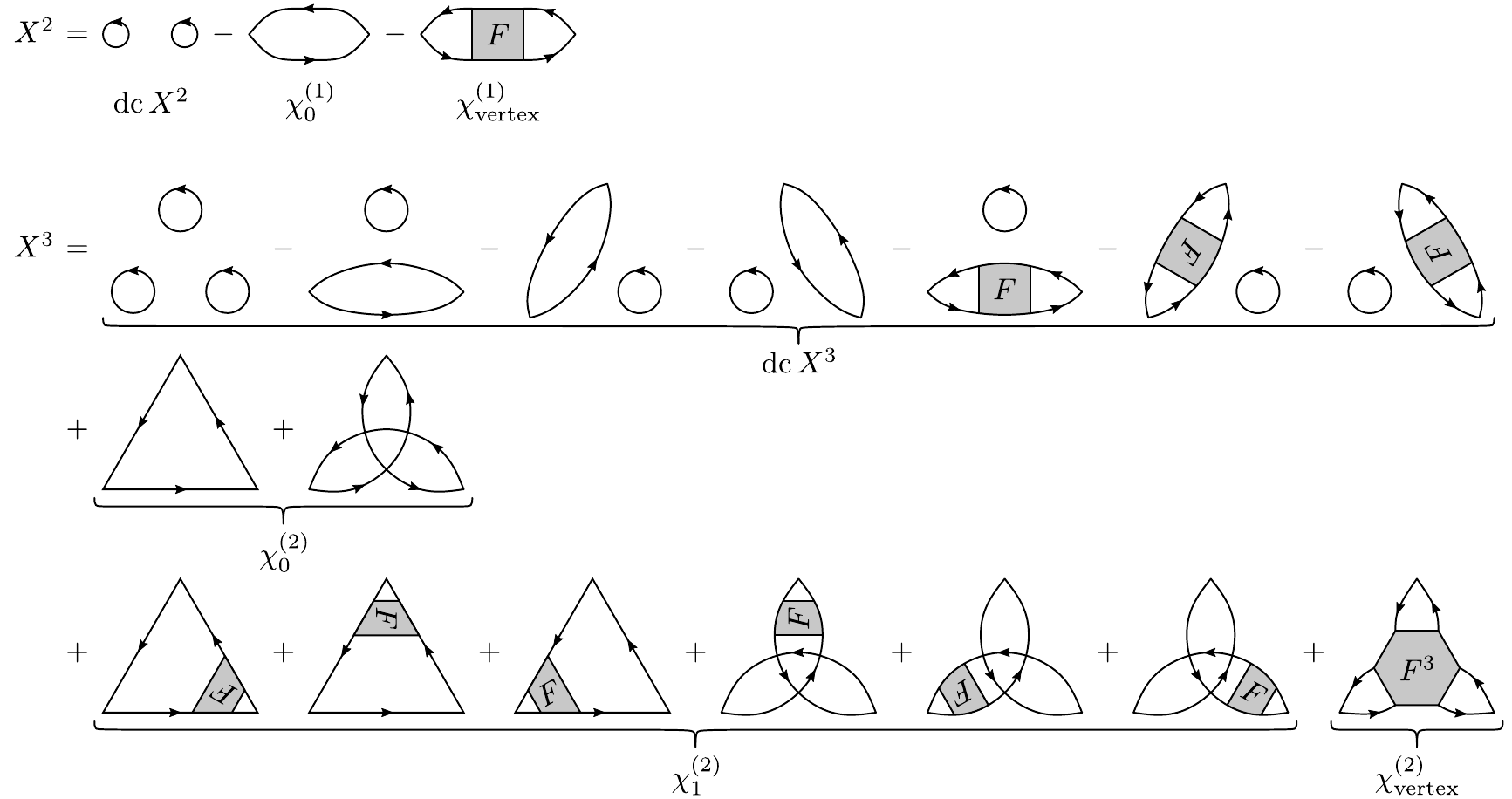}
    \caption{Diagrammatic representation of the decomposition of the full, two- and
             three-particle correlators $\corr$ into disconnected terms $\dc \corr$, bare
             or bubble terms $\chi_0$, first-order terms $\chi_1$ and full vertex terms
             $\chivertex$. Time and frequency labels are omitted to avoid clutter. They
             can easily be inferred from \cref{fig:correlators}.}
    \label{fig:decomposition}
\end{figure*}

To get a diagrammatic representation of the response functions we do a decomposition of
the full correlators and therefore Green's functions. For the three-particle case details
on this are found in \cref{sec:decomposition-of-g3}. The diagrammatic results are shown in
\cref{fig:decomposition}.
Here we see that the terms in \cref{eq:chi,eq:chi-2,eq:chi-omega,eq:chi-2-omega}, which we
already called disconnected, are in fact represented by disjoint diagrams.
\Cref{fig:decomposition} further decomposes the connected diagrams of $\chione$ and
$\chitwo$ and introduces the bare or bubble terms $\chi_0$, which only contain Green's
functions, the first-order terms $\chitwo_1$, which are three-particle diagrams with a
single two-particle vertex $F$, and $\chivertex$, which contains corrections from the
\enquote{largest} possible vertex. When studying the results in \cref{sec:results} we are
especially interested in $\chitwovertex$ since it contains diagrams where all three
particles interact with each other.
     \section{Models}
\label{sec:models}

For simplicity, we mainly employ the Anderson impurity model, but some results for the
Hubbard model as well as for the atomic limit, which is the same for both, are also shown.
The results for these three cases are presented in the corresponding subsections of
\cref{sec:results}.

\subsection{Atomic limit}
\label{ssec:al}

As a simple toy model, we consider the atomic limit with Hamiltonian $\HH_\text{AL} =
\epsilon (\nn_\up + \nn_\down) - h (\nn_\up - \nn_\down) + U \nn_\up \nn_\down$ where
$\epsilon=-U/2$ and $h$ is a magnetic field. This model can be solved exactly using the
Lehmann representation.

\subsection{Anderson impurity model}
\label{ssec:aim}

We use the Anderson impurity model (AIM) with (i) a single bath site and (ii) a constant
density of states (DOS). The former is chosen because it can easily be solved with exact
diagonalization and therefore serves as a test for the implementation of the
three-particle calculations. The latter is used for most other results because, while
still relatively simple and therefore fast to solve on modern computers, it already shows
effects of strong electronic correlation. The low computational complexity of the model is
important because of two things: First, we expect that the search for regions where
three-particle effects are relevant involves sampling a potentially large amount of points
in the phase diagram. Second, we solve the AIM by means of a quantum Monte Carlo (QMC)
solver. Once the interesting points are found, getting accurate, low-noise results for
$\chitwo$ can require a lot of QMC samples since the disconnected parts that have to be
subtracted first (see \cref{sec:theory}) potentially make up most of the correlation
function.

The Hamiltonian for the AIM is in general given by
\begin{equation}
    \begin{split}
        \HH_\text{AIM} = {}
        & \epsilon \nn + U \nn_\up \nn_\down
            + \sum_{k, \sigma} \epsilon^{\vphantom{\dagger}}_k
            \cdag_{k \sigma} \cc_{k \sigma} \\
        & {} + \sum_{k, \sigma} \pqty{
            V^{\vphantom{\dagger}}_k \fdag_{\sigma\vphantom{k}} \cc_{k \sigma}
            + V^{*\vphantom{\dagger}}_k \cdag_{k \sigma}
            \ff_{\sigma \vphantom{k}}},
    \end{split}
    \label{eq:aim-hamiltonian}
\end{equation}
where $\epsilon$, $\fdag_\sigma$ and $\ff_\sigma$ are the energy level as well as the
creation and annihilation operator of the impurity, $\nn^{\vphantom{\dagger}}_\sigma =
\fdag_\sigma \ff_\sigma$ is the impurity density operator for spin $\sigma$, $\nn =
\nn_\up + \nn_\down$ is the total density operator for the impurity, $U$ is the on-site
Coulomb interaction, $\epsilon^{\vphantom{\dagger}}_k$, $\cdag_{k, \sigma}$ and $\cc_{k
\sigma}$ are the energy levels as well as creation and annihilation operators for the
bath, and $V$ is the hybridization.

In the simplest case of only a single bath site, we have $\epsilon_{k = 1} = \epsilon_1$.
The second case we consider has a constant DOS
\begin{equation}
    \rho(\epsilon) =
    \begin{cases}
        \rho_0, & -D \leq \epsilon \leq D \\
        0, & \text{otherwise}
    \end{cases}
\end{equation}
and a real hybridization $V_k \equiv V$. This means that there is a continuous set of bath
sites with energies between $-D$ and $D$.

\subsection{Hubbard Model}
\label{ssec:hubbard-model}

The Hubbard model is one of the standard models when it comes to investigating strong
electronic correlations, but it is more difficult to solve than the AIM. For our
calculations we choose the single-band square-lattice Hubbard model which is defined by
the following Hamiltonian:
\begin{equation}
    \HH_\text{HM} = -t \sum_{\langle ij \rangle, \sigma} \pqty{
        \cdag_{i \sigma} \cc_{j \sigma} + \cdag_{j \sigma} \cc_{i \sigma}}
    + U \sum_i \nn^{\vphantom{\dagger}}_{i \up} \nn^{\vphantom{\dagger}}_{i \down}.
    \label{eq:hubbard-model-hamiltonian}
\end{equation}
As usual, $t$ is the hopping integral, $\sum_{\langle ij \rangle}$ denotes the sum over
nearest neighbors, $\cdag_{i \sigma}$ and $\cc_{i \sigma}$ are the creation and
annihilation operators for electrons with spin $\sigma$ at site $i$, $U$ is the on-site
interaction strength and $\nn^{\vphantom{\dagger}}_{i \sigma} = \cdag_{i \sigma} \cc_{i
\sigma}$ is the density operator for electrons with spin $\sigma$ at site $i$.

Since, the Hubbard model cannot be solved, also not numerically, except for very small
clusters, we employ DMFT \cite{Metzner1989,Georges1992,Jarrell1992,Georges1996} for an
approximate solution. DMFT actually maps the Hubbard model onto a self-consistent solution
of the AIM. The susceptibilities that we calculate here are local impurity
susceptibilities only. This means that these susceptibilities are actually also obtained
from the AIM, but now at DMFT self-consistency. They differ from the lattice
susceptibilities, also the local ones, since the applied fields can also affect the DMFT
bath of the auxiliary Anderson model. This effect is not taken into account here.
     \section{Results}
\label{sec:results}

The numerical results in this section are obtained with
\program{w2dynamics}~\cite{w2dynamics}, a continuous-time quantum Monte Carlo (CT-QMC)
solver using the hybridization expansion~\cite{Gull2011}. Only for the AIM with one bath
site we also employ exact diagonalization (ED). Let us also mention that numerical
renormalization group (NRG) has been successfully employed recently for calculating
multipoint correlators of the AIM~\cite{Lee2021}. Post-processing of the CT-QMC results is
done with the Python package \program{w2diag}~\cite{w2diag} written to, among other
things, implement the equations in \cref{sec:theory} and compute the first- and
second-order susceptibilities $\chione$ and $\chitwo$, involving two- and three-particle
correlators, respectively.

The results for the atomic limit are computed analytically through the Lehmann
representation.

A dataset containing all numerical data and plot scripts used to generate the figures in
this section is publicly available on the TU Wien Research Data repository \cite{dataset}.
The dataset also contains auxiliary data files, parameter files and submission scripts for
better reproducibility.

\subsection{Atomic limit}
\label{ssec:al_results}

In the atomic limit we only have four states: empty site, single occupation with spin
$\sigma$ $\up$ or $\down$, and double occupation with energies zero, $\epsilon \mp h$, and
$2 \epsilon + U$, respectively. We calculate the three-particle correlators of the atomic
limit, as defined in \cref{eq:3p-spin-corr,eq:3p-corr,eq:chiomom}, employing the Lehmann
representation in \cref{sec:lehmann}. As shown in \cref{sec:symmetries-of-3p-corr}, with
SU(2) symmetry and swapping relations we only obtain three independent flavor
combinations, $\alpha \beta \gamma = nnn, nzz, xyz$, for the second-order (three-operator)
susceptibility. Moreover, the first two flavor combinations correspond, in the atomic
limit, to conserved and mutually commuting operators. As a result, these three-particle
correlators are purely thermal: $\corr_{nnn}^{\omega_1 \omega_2} = \corr_{nnn}
\delta_{\omega_1 0} \delta_{\omega_2 0}$, $\corr_{nzz}^{\omega_1 \omega_2} =
\corr_{nzz}\delta_{\omega_1 0} \delta_{\omega_2 0}$. Only $\corr_{xyz}^{\omega_1
\omega_2}$ has a frequency structure.

Let us first consider the noninteracting case ($U=0$) at half-filling ($\epsilon=0$),
without a magnetic field ($h=0$). We evaluate the three-particle correlator using Wick's
theorem:
\begin{multline}
    \corr_{\sigma_1 \cdots \sigma_6}^{\omega_1 \omega_2} \\
    \begin{aligned}
        {} = {}
        & \beta^2 \ev{\nn_{\sigma_1}} \ev{\nn_{\sigma_2}} \ev{\nn_{\sigma_3}} \,
            \delta_{\sigma_1 \sigma_1'} \delta_{\sigma_2 \sigma_2'}
            \delta_{\sigma_3 \sigma_3'} \delta_{\omega_1 0} \delta_{\omega_2 0} \\
        & {} - \beta \! \ev{\nn_{\sigma_1}} \! \frac{1}{\beta} \!
            \sum_\nu G_{\sigma_2}^{\nu} G_{\sigma_3}^{\nu + \omega}
            \delta_{\sigma_1 \sigma_1'} \delta_{\sigma_2 \sigma_3'}
            \delta_{\sigma_3 \sigma_2'} \delta_{\omega_1 0} \\
        & {} - \beta \! \ev{\nn_{\sigma_2}} \! \frac{1}{\beta} \!
            \sum_\nu G_{\sigma_1}^{\nu} G_{\sigma_3}^{\nu + \omega}
            \delta_{\sigma_1 \sigma_3'} \delta_{\sigma_2 \sigma_2'}
            \delta_{\sigma_3 \sigma_1'} \delta_{\omega_2 0} \\
        & {} - \beta \! \ev{\nn_{\sigma_3}} \! \frac{1}{\beta} \!
            \sum_\nu G_{\sigma_1}^{\nu} G_{\sigma_2}^{\nu + \omega}
            \delta_{\sigma_1 \sigma_2'} \delta_{\sigma_2 \sigma_1'}
            \delta_{\sigma_3 \sigma_3'} \delta_{\omega_1, -\omega_2} \\
        & {} + \frac{1}{\beta} \! \sum_\nu G_{\sigma_1}^{\nu}
            G_{\sigma_2}^{\nu + \omega_1} G_{\sigma_3}^{\nu + \omega_1 + \omega_2} \\
        & {} + \frac{1}{\beta} \! \sum_\nu G_{\sigma_2}^{\nu}
            G_{\sigma_1}^{\nu + \omega_2} G_{\sigma_3}^{\nu + \omega_1 + \omega_2}.
    \end{aligned}
    \label{eq:x3wick}
\end{multline}
The first term with three densities $n$ corresponds to the fully disconnected, first
diagram for $\corr^3$ in \cref{fig:decomposition}. The terms two to four have one density
and a bare bubble susceptibility [for $U=0$: $\chi_d = \chi_m = -\frac{1}{\beta} \sum_\nu
G^{\nu} G^{\nu + \omega}$], as the next three diagrams for $\corr^3$ in
\cref{fig:decomposition}. All of these terms are disconnected and do not contribute to the
second-order, three-particle susceptibility.

The last two terms in \cref{eq:x3wick} are the bare bubble second-order susceptibility
$\chitwo_0$, represented diagrammatically in the second line of diagrams for $\corr^3$ in
\cref{fig:decomposition} [cf. \cref{eq:chi-nnn,eq:chi-nzz,eq:chi-xyz}]. This connected
part contains the essential three-particle information. While for $U=0$ it is given
through the (two) bare bubble diagrams, vertex corrections become important for $U\neq0$.
Namely, there are corrections with the two-particle vertex $F$ connecting two Green's
function lines as well as more complicated three-particle vertex corrections with $F^3$.
The latter connects all three Green's function lines of the bubble through interactions.

As a technical note: In \cref{eq:x3wick} the connected part appears next to the
disconnected term of order $\beta^2$. Therefore, a stochastic measurement of the
three-particle susceptibility has a less favorable signal-to-noise ratio at low
temperatures than the two-particle one, whose disconnected term is only of order $\beta$.

\begin{figure}
    \centering
    \includegraphics{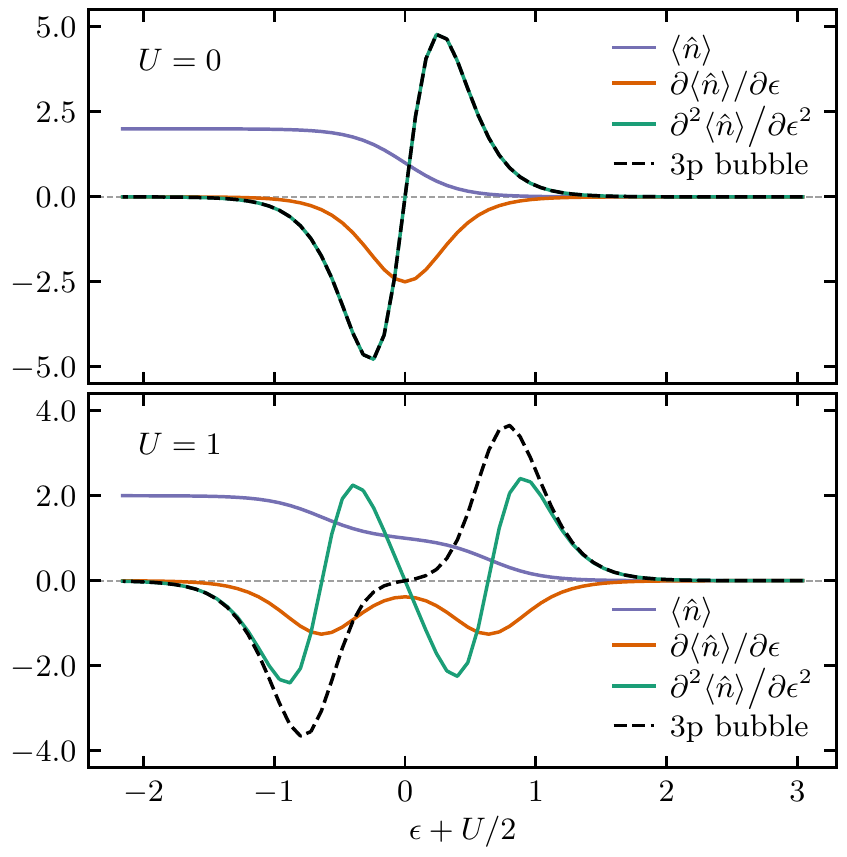}
    \caption{Density $\ev{\nn}$ as a function of the energy $\epsilon$ (purple) in the
    atomic limit ($\beta=5$, $h=0$). Orange and green curves show the first and second
    derivative, respectively; a black dashed line the three-particle bubble. Top panel:
    Noninteracting limit. Bottom panel: $U=1$.}
    \label{fig:al_deriv}
\end{figure}

The upper panel of \cref{fig:al_deriv} shows the density $\ev{\nn}$ of the noninteracting
system as a function of the energy $\epsilon$ ($\beta=5$), together with its first and
second derivatives with respect to $\epsilon$.
The latter is computed from \cref{eq:chi-2-omega}, but with partial instead of functional
derivatives and therefore static correlators, yielding
\begin{equation}
    \pdv[2]{\! \ev{\nn}}{\epsilon} =
        \corr^{00}_{nnn} + 3 \beta \! \ev{\nn} \pdv{\! \ev{\nn}}{\epsilon}
        - \beta^2 \ev{\nn} \ev{\nn} \ev{\nn},
    \label{eq:d2ndeps2}
\end{equation}
where $\pdv*{\! \ev{\nn}}{\epsilon} = -\corr^{0}_{nn} + \beta \ev{\nn} \ev{\nn}$ [cf.
\cref{eq:chi-d}] and $\ev{\nn}$ are computed in the usual way, see, e.g.,
Ref.~\cite{Medvedeva17}. We verified that \cref{eq:d2ndeps2} coincides with the analytical
and numerical second derivative of $\ev{\nn}$. Also drawn is the three-particle bubble
(dashed), which coincides exactly with the second derivative, as expected ($U=0$). For
negative (positive) $\epsilon$, double occupations (empty sites) are favorable. Hence, the
density $\ev{\nn}$ shows a crossover of width $1 / \beta$, and the other quantities follow
as derivatives.

Next, we turn on the interaction ($U=1$), which lifts the degeneracy of the empty, singly,
and doubly occupied states at $\epsilon=0$. As a result, the derivatives of $\ev{\nn}$ in
the lower panel of \cref{fig:al_deriv} acquire additional minima and maxima. It is
interesting to compare $\pdv*[2]{\! \ev{\nn}}{\epsilon}$ to the three-particle bubble,
which lacks vertex corrections. For large dopings the two curves coincide, which can be
considered as a perturbative regime where interaction effects are small. However, in the
correlated regime, near half-filling the three-particle bubble fails qualitatively. In
particular, it is unable to describe the curvature of $\ev{\nn}$ for $\epsilon$ in between
the Hubbard bands.

\begin{figure}
    \centering
    \includegraphics{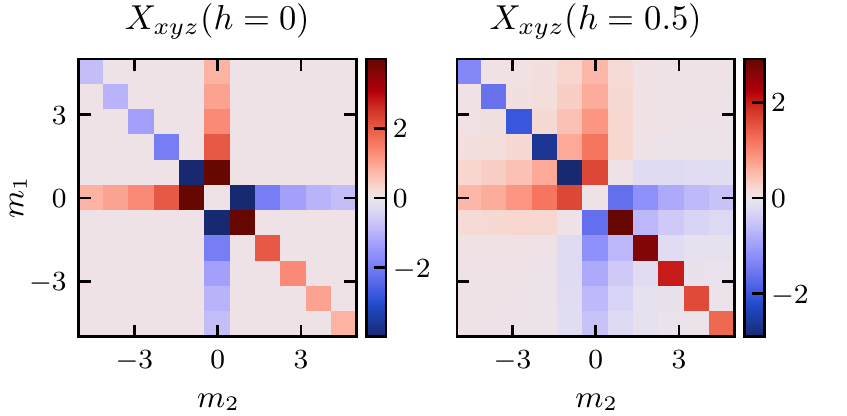}
    \caption{Full three-particle correlator with flavors $x,y,z$ drawn as a function of
             the two indices $m_{i=1,2}$ of the bosonic Matsubara frequencies $\omega_i =
             m_i 2 \pi T$; $U=\epsilon=0, \beta=5$. In the atomic limit only this flavor
             combination retains a frequency structure due to non-commutativity of the
             spin operators. Left: no magnetic field. Right: magnetic field in
             $z$-direction, $h=0.5$.}
    \label{fig:al_xyz}
\end{figure}

Finally, we consider the only three-particle correlation function that retains a
nontrivial frequency structure in the atomic limit: the chiral susceptibility
$\corr_{xyz}^{\omega_1 \omega_2} = \chixyz^{\omega_1 \omega_2}$. The left panel of
\cref{fig:al_xyz} shows this function for $U=0$, $\epsilon=0$, $h=0$, $\beta=5$.
This picture does not change qualitatively when $U$ is turned on (not shown), which
underlines that the frequency structure of $\corr_{xyz}$ is a result of the
non-commutativity of the spin operators among each other, rather than due to a specific
interaction regime. Notice also that the function is singular, that is, it vanishes
exactly away from the cross and diagonal structures, since each component of the spin
operator is conserved. This property does not persist for a finite magnetic field $h=0.5$
in the $z$ direction (right panel), which softens the cross structure, since it does not
commute with the $x$ and $y$ components of the spin operator.

\subsection{AIM with one bath site}
\label{ssec:single-site-aim}

To test the correctness of the implementation of the second-order response functions in
\program{w2diag} the results are compared against solutions of an AIM with a single bath
site obtained via exact diagonalization (ED). More precisely the density $n$ and the
linear, magnetic response function $\chi_{m}$ are computed with ED and then numerically
differentiated. This yields the right-hand side of the following two formulas
\begin{align}
    \chinnn^{00} & = \pdv[2]{\epsilon} \ev{\nn},
    \label{eq:chi-nnn-static}
    \\
    \chinzz^{00} & = -\pdv{\epsilon} \chi_{m},
    \label{eq:chi-nzz-static}
\end{align}
which are basically \cref{eq:chi-nnn,eq:chi-nzz} but with partial instead of functional
derivatives for denoting static response functions $\chinnn^{00}$ and $\chinzz^{00}$. The
left-hand side, is computed with \program{w2diag} from QMC results obtained with
\program{w2dynamcis}.

\begin{figure}
    \centering
    \includegraphics{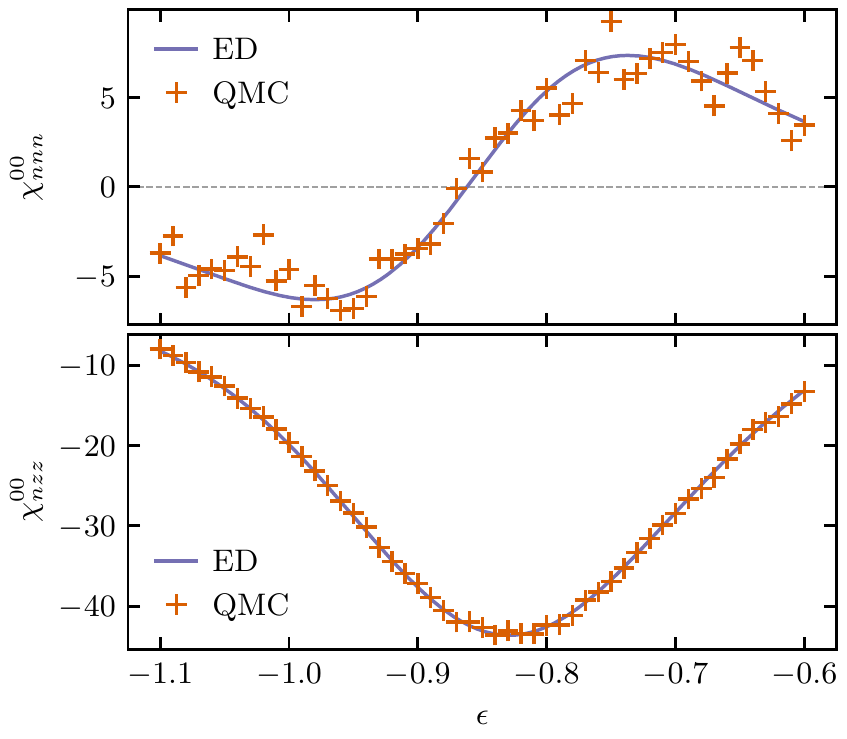}
    \caption{Comparison of exact (ED) and stochastic (QMC) results for the AIM at $U=1$,
    $\beta=20$, hybridizing with $V=0.2$ to a single bath site with energy
    $\epsilon_1=0.25$. The top plot shows the static, second-order, density response
    function $\chinnn^{00}$, while the bottom one shows the static, second-order,
    density-magnetic response function $\chinzz^{00}$.}
    \label{fig:single-site-aim-v0.2}
\end{figure}

\begin{figure}
    \centering
    \includegraphics{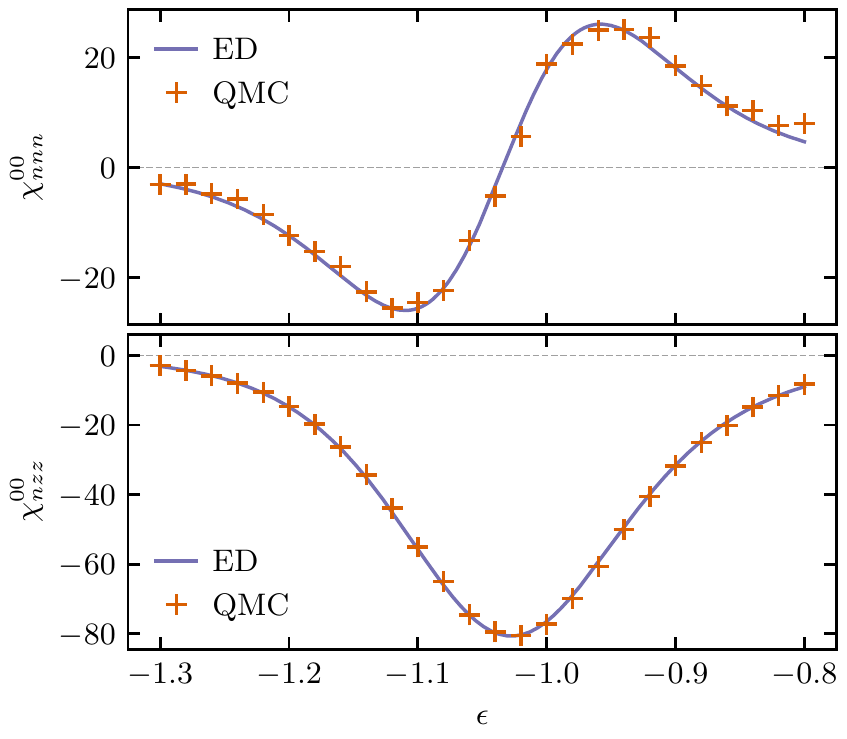}
    \caption{Comparison of exact (ED) and stochastic (QMC) results for the AIM with a
    single bath site. Same as \cref{fig:single-site-aim-v0.2}, but now at $U=1$,
    $\beta=20$, $V=0.05$ and $\epsilon_1=0$.}
    \label{fig:single-site-aim-v0.05}
\end{figure}

The results for $U=1$, $\beta=20$, $V=0.2$ and $\epsilon_1=0.25$ are shown as a function
of $\epsilon$ in \cref{fig:single-site-aim-v0.2}; those for $V=0.05$ and $\epsilon_1=0$
with the same $U$ and $\beta$ are plotted in \cref{fig:single-site-aim-v0.05}.
We see that the agreement between stochastic and exact results is very good except for
$\chinnn$ in \cref{fig:single-site-aim-v0.2} where the large noise prevents precise
statements.

This trend of higher noise in the data for $\chinnn$ is something we observe in almost
every computation and can be explained as follows: First, we measure the full
three-particle correlators with similar relative noise but since at least the static
component $\corr^{00}$ is usually larger in the $nnn$ channel than in the $nzz$ channel,
the absolute error is also larger there. Second, and more importantly, when looking at
\cref{eq:chi-2-omega} we see that for $\chinnn$ we have to subtract all four disconnected
terms from $\corr_{nnn}$, while for $\chinzz$ three of the four terms vanish because
$\ev{\zz} = 0$. Therefore, the magnitude of $\chinnn^{00}$ is often smaller than that of
$\chinzz^{00}$. Together this explains why the results for $\chinnn^{00}$ can have
significantly higher relative noise than those for $\chinzz^{00}$.
\Cref{fig:const-dos-aim-of-epsilon} in the next subsection shows this most dramatically.

\subsection{AIM with constant DOS}
\label{ssec:constant-dos-aim}

To find an area with potentially large, second-order effects we do calculations at two to
five times the Kondo temperature and make an $\epsilon$-scan starting from $-U/2$ going to
smaller values. The idea behind this is to find larger non-linear dynamics; and, going
away from particle-hole symmetry $\epsilon=-U/2$ reduces the Kondo temperature so that the
derivative with respect to $\epsilon$ should be sizable.

The chosen parameters are $D=10$, $U=6$, $V=2$ and $\beta=18$. They satisfy $D > U \gtrsim
V$ so according to \cite[p.~165ff]{Hewson1993} we can estimate the Kondo temperature to be
$1/T_K=\beta_K \approx 64$ which means that $\beta / \beta_K \approx 3.6$.

\begin{figure}
    \centering
    \includegraphics{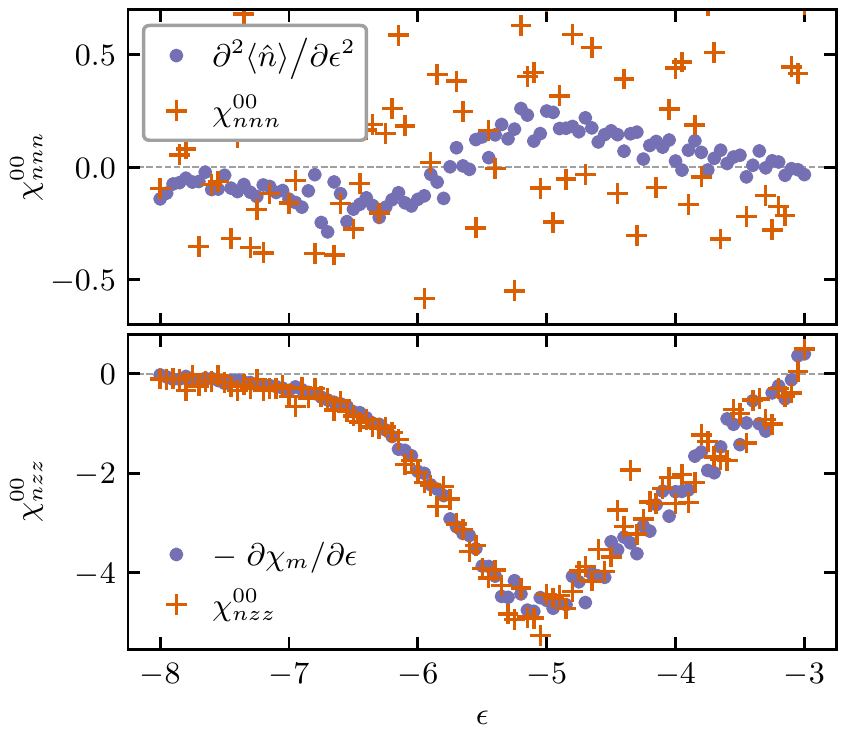}
    \caption{Scan of $\pdv*[2]{\! \ev{\nn}}{\epsilon}$, $-\pdv*{\chi_m}{\epsilon}$,
             $\chinnn$ and $\chinzz$ vs.~$\epsilon$ for an AIM with constant DOS at
             $D=10$, $U=6$, $V=2$ and $\beta=18$. The noise of $\chinnn$ is even larger
             than depicted. Its data points are actually outside the plotted region and
             between $-1.5$ and 1.5. However, with that range on the $y$-axis the extrema
             of $\pdv*[2]{\! \ev{\nn}}{\epsilon}$ would hardly be noticeable.}
    \label{fig:const-dos-aim-of-epsilon}
\end{figure}

\Cref{fig:const-dos-aim-of-epsilon} shows the second derivative of the density
$\pdv*[2]{\! \ev{\nn}}{\epsilon}$, the first derivative of the linear, magnetic response
function $\pdv*{\chi_m}{\epsilon}$ as well as the static, second-order, density and
density-magnetic response functions $\chinnn^{00}$ and $\chinzz^{00}$ plotted over
$\epsilon$.
This time $\chinnn^{00}$ is so noisy that no useful information can be extracted (see the
discussion at the end of \cref{ssec:single-site-aim} for an explanation). Nevertheless, we
clearly observe the largest, second-order effects around $\epsilon = -5$ and also
$\epsilon = -6.6$, so we choose those points for closer, frequency-resolved investigation.

\begin{figure}
    \centering
    \includegraphics{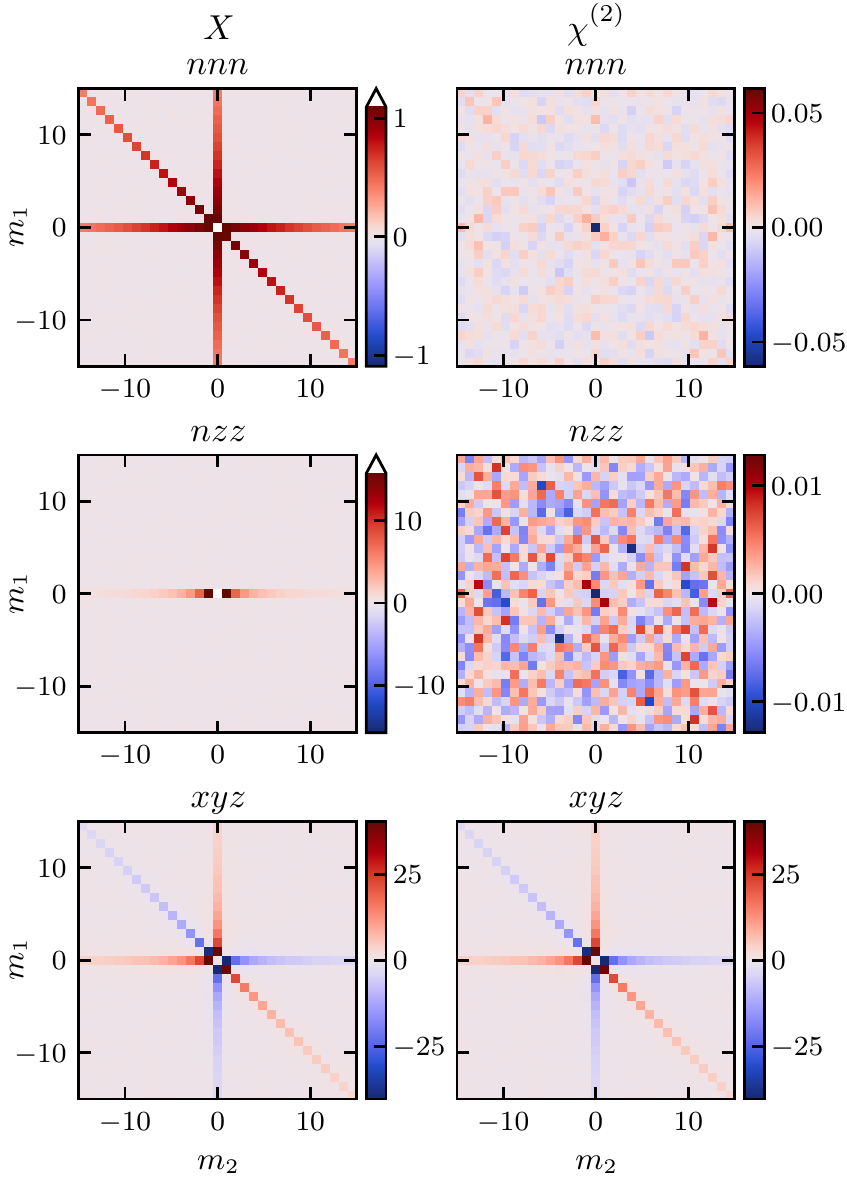}
    \caption{Full correlation functions $\corr$ (left column) and second-order response
    functions $\chitwo$ (right column) vs.~the two indices $m_{i=1,2}$ of the bosonic
    Matsubara frequencies $\omega_i = m_i 2 \pi T$ in the density ($nnn$; top row),
    density-magnetic ($nzz$; center row), and chiral ($xyz$; bottom row) channel for an
    AIM with constant DOS at $D=10$, $U=6$, $V=2$, $\beta=18$ and half-filling, i.e.,
    $\epsilon = -3$ and $n=1$. The color bars for $\corr_{nnn}$ and $\corr_{nzz}$ exclude
    the largest value at the center because it would dominate the plots. These values are
    $\corr_{nnn}^{0 0} \approx 327$ and $\corr_{nzz}^{0 0} \approx 171$.}
    \label{fig:const-dos-aim-epsilon3}
\end{figure}

First, however, we take a look at half-filling, i.e, $\epsilon = -U/2 = -3$ and $n=1$.
\Cref{fig:const-dos-aim-epsilon3} shows the full correlation functions $\corr^{\omega_1
\omega_2}$ (left column) and the second-order response functions $\chi^{\omega_1
\omega_2}$ (right column) in the density, density-magnetic, and chiral channel.
The color bars for $\corr_{nnn}$ and $\corr_{nzz}$ exclude the largest value at the center
because it would dominate the plots with values of $\corr_{nnn}^{0 0} \approx 327$ and
$\corr_{nzz}^{0 0} \approx 171$. As expected from the discussion at the end of
\cref{sec:theory} we see that, in this case, $\chinnn$ and $\chinzz$ vanish (the higher
absolute noise at the center of $\chinnn$ comes from the large value of $\corr_{nnn}^{0
0}$). There is also no difference between the full correlator and the connected parts in
the chiral channel. $\corr_{nnn}$ and $\corr_{nzz}$ clearly show the structure of the
disconnected parts with their $\delta_{\omega_i 0}$ terms [see
\cref{eq:chi-spin-123-omega,eq:chi-spin-123bar-omega,eq:chi-spin-1bar23bar-omega,eq:chi-spin-12bar3-omega}].
The latter are also responsible for the large values at $\omega_1 = \omega_2 = 0$ that are
clipped from the color bar. Although there are no disconnected terms for $\corr_{xyz}$, it
shows similar \enquote{cross}-like but antisymmetric structures. Additionally, note that
in \cref{fig:const-dos-aim-epsilon3} we see the following order when comparing the
magnitudes of the non-static parts: $\corr_{xyz} > \corr_{nzz} > \corr_{nnn}$. For the
static components this order is exactly reversed.

\begin{figure*}
    \centering
    \includegraphics{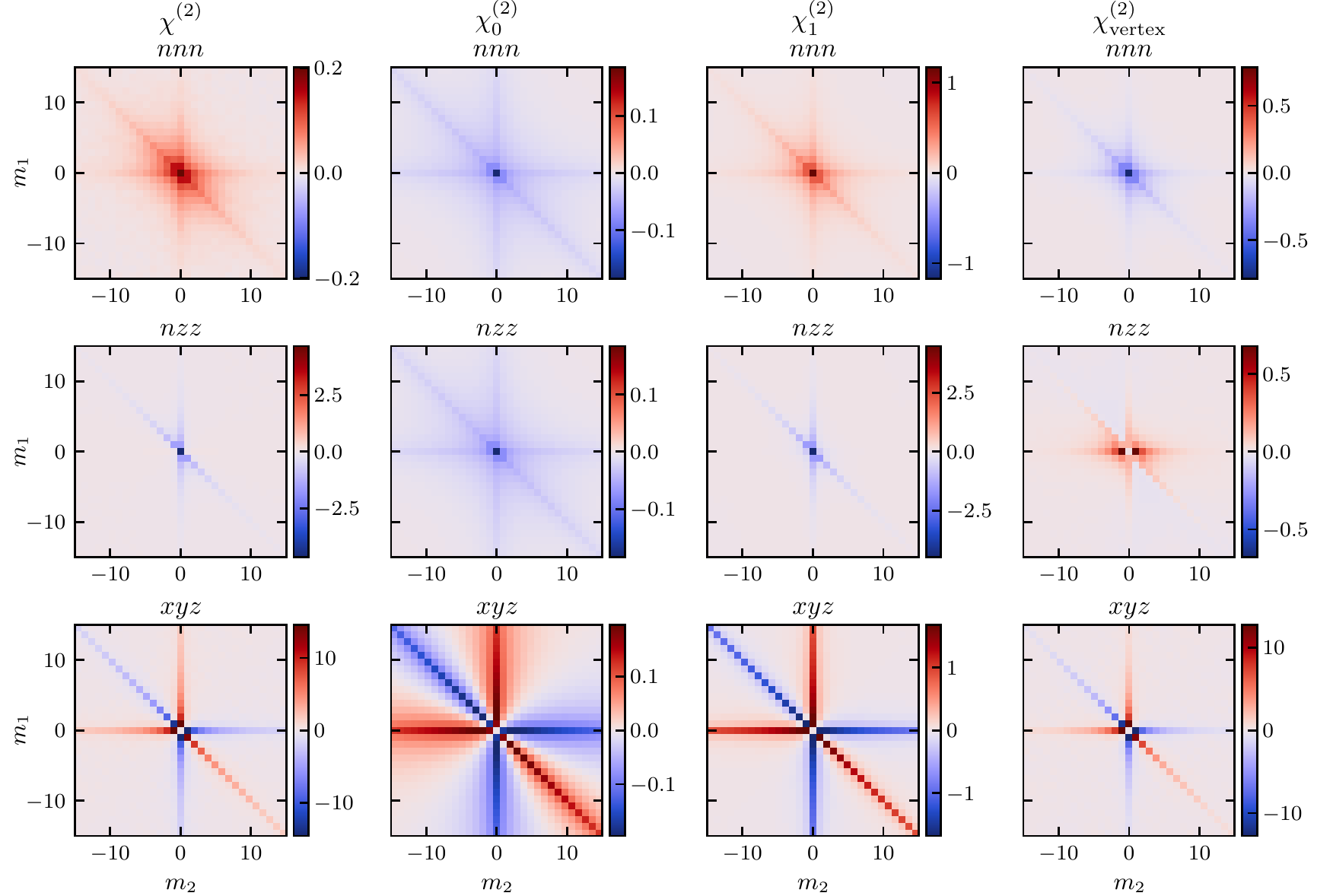}
    \caption{Second-order response functions $\chitwo$ and their decomposition into the
    bubble terms $\chitwo_0$, first-order terms $\chitwo_1$ and vertex terms
    $\chitwovertex$ in the density ($nnn$), density-magnetic ($nzz$), and chiral ($xyz$)
    channel for an AIM with constant DOS at $D=10$, $U=6$, $V=2$, $\beta=18$ and $\epsilon
    = -5$ corresponding to $n \approx 1.22$.}
    \label{fig:const-dos-aim-epsilon5}
\end{figure*}

As discussed above, the results for the largest, second-order effects in the density and
density-magnetic channel are found approximately at $\epsilon = -5$ in
\Cref{fig:const-dos-aim-of-epsilon}, corresponding to $n \approx 1.22$. They are shown in
\cref{fig:const-dos-aim-epsilon5} where we plot the full, second-order response function
$\chitwo$, as well as its decomposition into the bubble terms $\chitwo_0$, first-order
terms $\chitwo_1$, and vertex terms $\chitwovertex$ in all three channels.
For the density-like response functions, i.e., $\chinnn$ and $\chinzz$, even after
subtracting the disconnected terms we still see the maximum at the center point and
\enquote{cross}-like structures along the $\omega_i = 0$ lines. Those features are,
however, much less pronounced and more washed out when compared to those of $\corr$ in
\cref{fig:const-dos-aim-epsilon3}. Since there are generally no disconnected terms in the
chiral channel ($\chixyz=\corr_{xyz}$) the plot of $\chixyz$ looks almost exactly the same
as before. Compared to \cref{fig:const-dos-aim-epsilon3} only the magnitude is reduced
because of the different doping. When looking at the decomposition, the density-like
channels all look rather similar and soft while the features in the chiral channel are
much more pronounced and long-ranged. Regardless of that, the bubble terms $\chitwo_0$ are
of similar magnitude for all channels but never a good approximation for the whole
second-order response functions. They are too small and in the density channel the bubble
even has the wrong sign. The first order terms $\chitwo_1$ are larger (sometimes too
large) and always have the right sign, but that is still not enough. Across all three
channels the corrections from the three-particle vertex $\chitwovertex$ have sizable
contributions that cannot be neglected. Especially $\chixyz$ is dominated by these terms.
When comparing maximum magnitudes of the second-order response functions we see the same
ordering as for the non-static parts of $\corr$ in the half-filled case: $\chixyz >
\chinzz > \chinnn$.

\begin{figure*}
    \centering
    \includegraphics{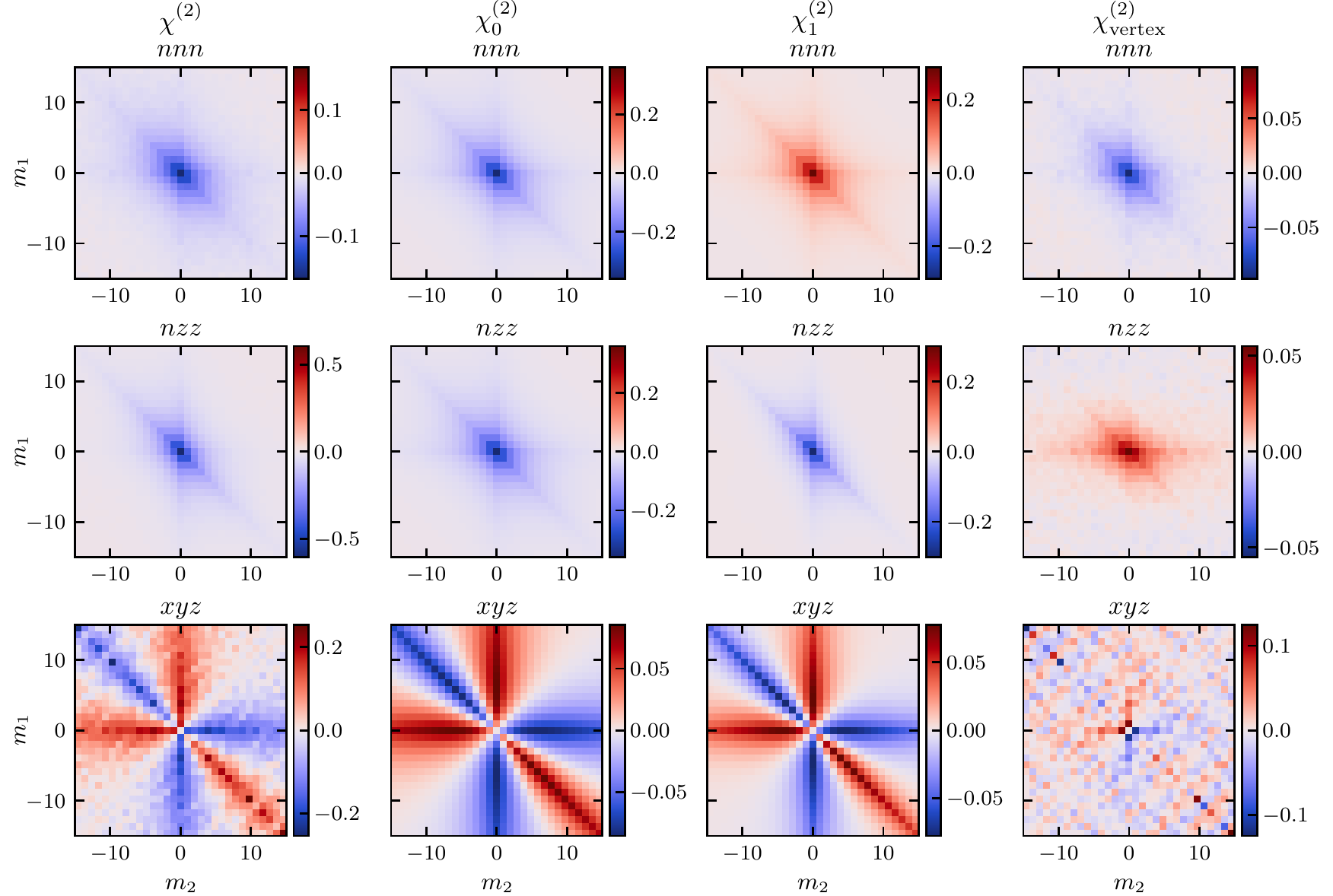}
    \caption{Same as \cref{fig:const-dos-aim-epsilon5} except now $\epsilon = -6.6$
    corresponding to $n \approx 1.68$.}
    \label{fig:const-dos-aim-epsilon6.6}
\end{figure*}

\Cref{fig:const-dos-aim-epsilon6.6} shows the same plots as
\cref{fig:const-dos-aim-epsilon5} but for $\epsilon = -6.6$, corresponding to $n \approx
1.68$, where the second-largest, second-order effects in the density channel are found.
We see in all plots that the features are much less pronounced. Especially the plots of
the chiral channel are more washed out and, since the magnitude of $\chixyz$ is much lower
than when closer to half-filling, noise becomes a problem, particularly for the
three-particle vertex corrections. The sign change of $\chinnn$ is expected when looking
at \cref{fig:const-dos-aim-of-epsilon} and means that the bubble now has the correct sign
across all channels. In general, we see that at this higher doping the bubble becomes a
better approximation while $\chitwo_1$ and $\chitwovertex$ become smaller.

Finally, we take a look at the asymptotic behavior of the second-order response functions
in the limit of large Matsubara frequencies $\omega$. More precisely we look at 1d cuts
along $\omega_1 = 0$ and $\omega_2 = \omega$. From the detailed derivation in
\cref{sec:asymptotic-behavior-of-chi-2} we get
\begin{align}
    \chinnn^{0 \omega} \approx {} & -\frac{1}{\omega^2} \pdv{\ev{H_V}}{\epsilon},
    \label{eq:chi-nnn-asymptote}
    \\
    \chinzz^{0 \omega} \approx {} & -\frac{1}{\omega^2} \pdv{\ev{H_V}}{\epsilon},
    \label{eq:chi-nzz-asymptote}
    \\
    \chixyz^{0 \omega} \approx {} & -\frac{2}{\omega} \chi_m^0,
    \label{eq:chi-xyz-asymptote}
\end{align}
where $\chi^0_m$ is the static, linear magnetic response function and $H_V$ is the
hybridization term in the Hamiltonian of the AIM [last term in \cref{eq:aim-hamiltonian}].
Its derivative reads
\begin{equation}
    -\pdv{\epsilon} \ev{H_V}
    = \frac{4}{\beta} \sum_\nu \Delta_\up^\nu \pqty{
        \Pthree_{\up\up}^{\nu 0} + \Pthree_{\up\down}^{\nu 0}
        + \beta (2 - \ev{\nn}) G_\up^\nu},
    \label{eq:dhv-depsilon}
\end{equation}
where $\Delta$ is the hybridization function, $\Pthree$ is the partially contracted
two-particle Green's function $\Pthree^{\nu' \omega} = \frac{1}{\beta} \sum_\nu G^{\nu
\nu' \omega}$, $G^\nu$ is the one-particle Green's function, and we use SU(2) symmetry.
Note that the $1 / \omega$ terms for the density and density-magnetic channel vanish
because they are proportional to $\comm{\nn}{\nn}$ and $\comm{\zz}{\zz}$, respectively.

\begin{figure}
    \centering
    \includegraphics{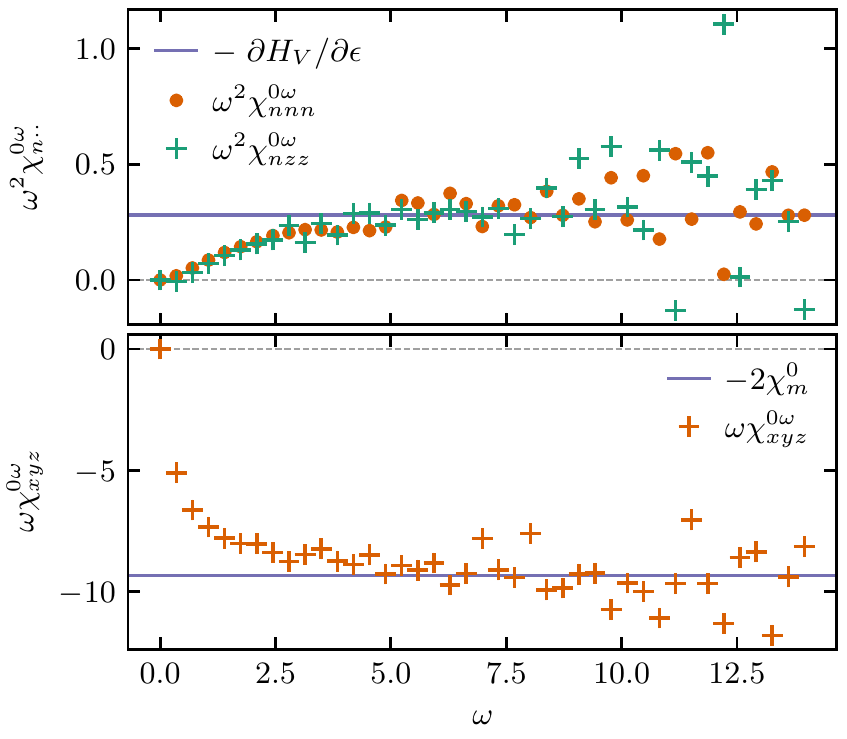}
    \caption{Analysis of the asymptotic behavior of the second-order response functions at
    $\omega_1 = 0$, $\omega_2 = \omega$. The top plot shows the density and
    density-magnetic channel multiplied by $\omega^2$, while in the bottom one the chiral
    channel is multiplied by $\omega$. The dots and pluses represent the numerical data
    computed for an AIM with constant DOS at $D=10$, $U=6$, $V=2$, $\beta=18$ and
    $\epsilon = -5$. These are the same parameters as in
    \cref{fig:const-dos-aim-epsilon5}. The solid lines are the analytically calculated
    asymptotic behavior taken from
    \cref{eq:chi-nnn-asymptote,eq:chi-nzz-asymptote,eq:chi-xyz-asymptote}.}
    \label{fig:const-dos-aim-asymptotes}
\end{figure}

\Cref{fig:const-dos-aim-asymptotes} compares the analytical results of
\cref{eq:chi-nnn-asymptote,eq:chi-nzz-asymptote,eq:chi-xyz-asymptote} with the numerical
data for $\chi^{0 \omega}$ at $\epsilon = -5$.
The equations are multiplied with $\omega^2$ and $\omega$, respectively, because the tails
drop so fast that a good comparison at medium to high frequencies would hardly be possible
otherwise. This, however, also amplifies the noise of the numerical results. Nevertheless,
we see a good agreement starting at frequencies as low as $\omega \approx 4$.

\subsection{Hubbard model}
\label{ssec:hubbard-model-results}

\begin{figure*}
    \centering
    \includegraphics{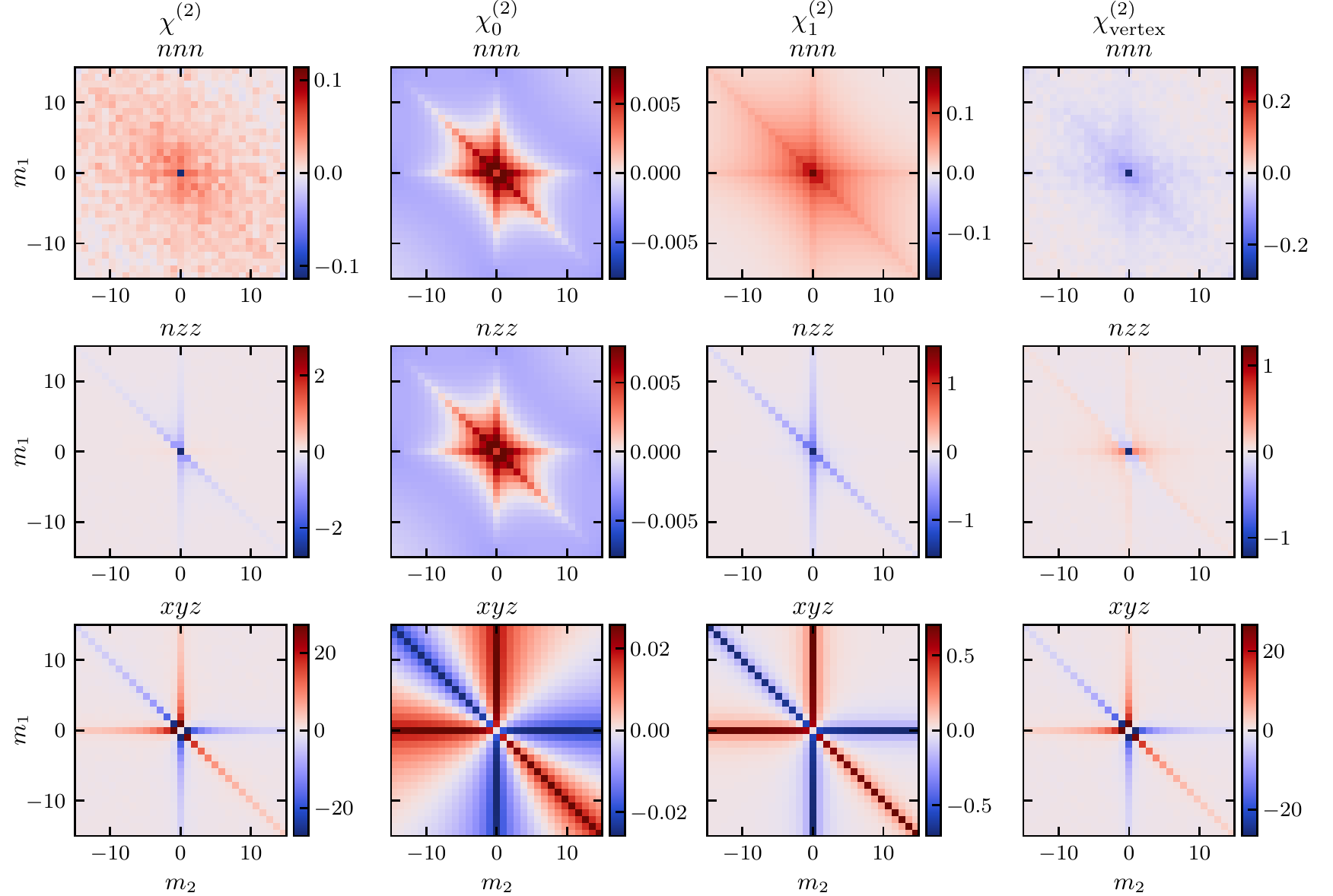}
    \caption{Second-order response functions $\chitwo$ and their decomposition into the
    bubble terms $\chitwo_0$, first-order terms $\chitwo_1$ and vertex terms
    $\chitwovertex$ in the density ($nnn$), density-magnetic ($nzz$), and chiral ($xyz$)
    channel for a single-band, square-lattice Hubbard model at $n=1.1$, $U=12$ and
    $\beta=20$.}
    \label{fig:hubbard-model}
\end{figure*}

\Cref{fig:hubbard-model} shows the same plots as
\cref{fig:const-dos-aim-epsilon5,fig:const-dos-aim-epsilon6.6} namely the full,
second-order response function $\chitwo$, the bubble terms $\chitwo_0$, the first-order
terms $\chitwo_1$ and the vertex terms $\chitwovertex$ in the density, density-magnetic,
and chiral channel.
This time, however, computed for the two-dimensional Hubbard model on a square lattice
with $n=1.1$, $t=1$ (i.e. $D=4$), $U=12$ and $\beta=20$.\footnote{Note that in this case
$\chitwo$ is only the local, second-order response as discussed above. The total one would
have to include the change in the hybridization function as well.} We see that for the
most part the results look similar to those for the AIM with constant DOS at $\epsilon =
-5$ (\cref{fig:const-dos-aim-epsilon5}). $\chinnn$ seems even more washed out, but the
biggest differences show the bubble terms of the density and density-magnetic channel
which are qualitatively different and no longer completely negative. They have a rather
steep positive hill centered around $\omega_1 = \omega_2 = 0$ and a slowly decaying
negative background at higher frequencies. The main takeaway that $\chitwo_0$ and
$\chitwo_1$ are bad approximations is, however, still valid.
     \section{Conclusion}
\label{sec:conclusion}

We have derived the equations for the frequency-resolved, non-linear response from
three-particle correlators. These are made up from three bosonic operators with three time
arguments (or two time differences or frequencies). We here focused on the local
correlator, the three-particle Green's function and susceptibility ($\chitwo$) on an
impurity site. However, the equations derived can be straight-forwardly extended to
non-local correlators adding a site index for each time index. We further showed how the
three-particle quantities are decomposed: This involves disconnected diagrams that do not
contribute to $\chitwo$ as well as two bubble diagrams $\chitwo_0$ without vertex
corrections, very similar as for the two-particle correlators. Then there are diagrams
$\chitwo_1$ consisting of a single particle propagator and two propagators connected by a
two-particle vertex as well as genuine three-particle vertex diagrams $\chitwovertex$ that
connect all incoming and outgoing lines (cf.~\cref{fig:decomposition}). The asymptotic
behavior of the correlators is given by a $1/\omega^2$ or $1/\omega$ term, depending on
whether the bosonic operators commute or not.

We have computed the correlators numerically using continuous-time QMC for the atomic
limit as well as for the AIM with a single-site, a flat DOS and at DMFT self-consistency
for the two-dimensional Hubbard model. We find a sizable non-linear density and
density-magnetic response functions at high doping. At half-filling, these two non-linear
responses vanish by symmetry. The pure density ($nnn$) response function suffers from the
fact that there are large contributions from disconnected terms that need to be
subtracted. This leads to a rather high level of numerical noise for the actual response
function. The chiral ($xyz$) response function is also sizable. It contributes at
half-filling and decreases with doping. It has the largest three-particle vertex
contributions. For all three non-vanishing local response functions ($nnn$, $nzz$, and
$xyz$) the three-particle vertex cannot be neglected for relevant ranges of the local
one-particle potential $\epsilon$.
     \section{Outlook}
\label{sec:outlook}

Physically the $nnn$ and $nzz$ response functions can, in the case of a static third
bosonic $n$ operator be related to the change of the charge susceptibility and magnetic
susceptibility with respect to a change of the local one-particle potential. If the third
bosonic operator becomes time or frequency dependent, we have the corresponding changes
against a dynamic one-particle potential. The $nnn$ susceptibility also describes the
non-linear charge response, and $nzz$ the non-linear charge response to an applied
magnetic field.

The chiral $xyz$ susceptibility is arguably the most exotic response as there is no
correspondence on the two-particle level; the $xy$, $xz$, and $yz$ susceptibilities all
vanish in the paramagnetic phase with SU(2) symmetry. The chiral $xyz$ susceptibility
describes a non-linear response of the spin in $x$-direction if (time-dependent) magnetic
fields in both $y$- and $z$-direction are applied. If one of the magnetic fields, say the
one in $z$-direction, was large and static it is akin a nuclear magnetic resonance
experiment. However, we are here in second-order response, i.e., only have a weak field in
$z$-direction. Nonetheless, there are ideas and efforts to actually measure this response
function~\cite{Alpicheshev2022}.

We have seen that the contribution of the three-particle vertex $F^3$ to the
three-particle susceptibility $\chitwo$ is in general not small, but comparable to (or
even larger than) the bare bubble contribution $\chitwo_0$ and contributions $\chitwo_1$
from two-particle vertices plus a disconnected propagator. This means that previous
approaches to calculate non-linear responses, such as the Hall and Raman response which
only included $\chitwo_0$ or at most $\chitwo_1$ need to be reassessed for strongly
correlated electron systems. Numerical approaches, as we have employed here for the local
susceptibilities, become prohibitively expensive for the full lattice. This calls for
developing approaches such as a three-particle Bethe--Salpeter equation for calculating
the three-particle vertex $F^3$ also in the non-local case.
     \begin{acknowledgments}
    We thank Z.~Alpicheshev, A.~Kauch, J.~Kaufmann, M.~Pickem, C.~Eckhardt and T.~Ribic
    for valuable and helpful discussions. This project has been funded by the Austrian
    Science Funds (FWF) through project P 32044. It also profited from discussions within
    the FWF SFB F86 Q-M\&S. Calculations have been done on the Vienna Scientific Cluster
    (VSC).
\end{acknowledgments}
 
    \appendix

    \section{Non-linear response theory (NLRT)}
\label{sec:nlrt}

\subsection{NLRT in real time and frequency}
\label{ssec:real-nlrt}


We assume that the system is coupled to external classical fields $F_j(t) \in \mathbb{R}$
and $\AA_j^\dagger = \AA_j$ as
\begin{equation}
    \HHH(t) = \HH - \sum_j \AA_j F_j(t).
    \label{eqn:Appendix:NLRTR:Hamiltonian}
\end{equation}
One example discussed in the main text is a spin $\AA_j=\qftop{S}_j$ coupled to a magnetic
field $F_j=B_j$ where $j=x,y,z$. We furthermore assume that the dynamics of the density
matrix $\qftop \rho$ are governed by the von Neumann equation:
\begin{equation}
    \frac{\dd}{\dd t} \qftop{\rho}(t) = - \ii [\HHH(t),\qftop{\rho}(t)],
    \label{eqn:Appendix:NLRTR:vonNeumann}
\end{equation}
with $\qftop{\rho}(t_0)=\frac{1}{Z} \ee^{-\beta \HH} =: \qftop{\rho}_0$. In the
interaction picture (with respect to $\HH$) the solution to the von Neumann equation can
be written as an infinite series of the following form:
\begin{equation}
    \begin{split}
        \qftop{\rho}(t) = {} & \qftop{\rho}_0 + \ii \sum_j \int_{t_0}^{t} \dd t' \, [\AA_j(t'),\qftop{\rho}(t')] F_j(t')\\
        = {} & \qftop{\rho}_0 \; \\
        & + \sum_{\ell=1}^{\infty} \sum_{j_1, j_2, \dots, j_\ell}
        \int_{t_0}^{t} \dd t_1 \int_{t_0}^{t_1} \dd t_2 \dots \int_{t_0}^{t_{\ell-1}} \dd t_\ell \\
        & \quad [\AA_{j_1}(t_1), [\AA_{j_2}(t_2), \dots, [\AA_{j_\ell}(t_\ell), \qftop\rho_0]] \dots]\\
        & \quad F_{j_1}(t_1) F_{j_2}(t_2) \dots F_{j_\ell}(t_\ell)
    \end{split}
    \label{eqn:Appendix:NLRTR:rho}
\end{equation}
The expectation value of a generic operator
$\ev*{\AA_{i}(t)}_{\vect{F}}=\Tr[\qftop{\rho}(t) \AA_i(t)]$ can now be evaluated in a
straightforward way. It is convenient to use a matrix identity:
\begin{multline}
    \Tr([\AA_1, [\AA_2, \dots [\AA_\ell, \qftop{\rho}_0] \dots ]] \, \AA_i) \\
    = \Tr([[\dots[\AA_i, \AA_1], \AA_2], \dots, \AA_\ell] \, \qftop{\rho}_0).
    \label{eqn:Appendix:NLRTR:CommutatorIdentity}
\end{multline}
This allows us to express the expectation value $\ev*{\AA_{i}(t)}$ as
\begin{equation}
    \begin{array}{rl}
        \ev*{\AA_{i}(t)}_{\vect{F}} =& \ev*{\AA}_{\vect{F}=0} + \sum_{\ell=1}^{\infty} \ev*{\Delta \AA_i(t)}^{(\ell)}\\
        :=&\ev*{\AA}_{\vect{F}=0} +
        \sum_{\ell=1}^{\infty} \sum_{j_1 j_2 \dots j_\ell} \times
        \\ &
        \int_{-\infty}^{\infty} \dd t_1 \int_{-\infty}^{\infty} \dd t_2 \dots \int_{-\infty}^{\infty} \dd t_\ell
        \\ &
        \:\: F_{j_1}(t_1) F_{j_2}(t_2) \dots F_{j_\ell}(t_\ell) \\&
        \:\: \chi^R_{j_1 j_2 \dots j_\ell i}(t_1, t_2, \dots, t_\ell, t),
    \end{array}
    \label{eqn:Appendix:NLRTR:A}
\end{equation}
where the (generalized) Kubo susceptibility is
\begin{multline}
    \chi^R_{j_1 j_2 \dots j_\ell i}(t_1, t_2, \dots, t_\ell, t) \\
    \begin{aligned}
        = {} & \ii^\ell \; \theta_{t > t_1 > t_2 > \dots > t_\ell > t_0} \\
        & \times \! \ev*{[[\dots [\AA_i(t), \AA_{j_1}(t_1)], \AA_{j_2}(t_2)],
                          \dots, \AA_{j_\ell}(t_\ell)]}.
    \end{aligned}
    \label{eqn:Appendix:NLRTR:GeneralizedKuboSusceptibility}
\end{multline}
Kubo already remarked in \cite{Kubo1957}[Eqs.~2.28--2.29] that his formalism is not
limited to linear response theory. Nonetheless, we call the quantity in
\cref{eqn:Appendix:NLRTR:GeneralizedKuboSusceptibility} generalized Kubo susceptibility
(although the generalization was done already by Kubo himself in the original work).

For the Fourier transform it is convenient to assume $F_i(t\!<\!t_0)\!=\!0$, i.e., that
the field is only switched on for positive times. The Fourier transform of a contribution
of order $\ell$ in the external field $F$ enjoys the representation given in
\cref{eqn:Appendix:NLRTR:InOmega},
\begin{equation}
    \begin{array}{rcl}
        \ev*{\Delta \AA_i(\omega)}^{(\ell)} &=&
        \frac{1}{(2\pi)^{\ell-1}}
        \int_{-\infty}^{\infty} \dd\omega_1 \int_{-\infty}^{\infty} \dd\omega_2 \dots \int_{-\infty}^{\infty} \dd\omega_\ell \\
        && \delta(\omega - \sum_{i=1}^\ell \omega_{i}) \\
        && F_{j_1}(\omega_1) F_{j_2}(\omega_2) \dots F_{j_\ell}(\omega_\ell) \\
        && \chi^{R}_{j_1 j_2 \dots j_\ell i}(-\omega_1, -\omega_2, \dots, -\omega_\ell, t=0),
    \end{array}
    \label{eqn:Appendix:NLRTR:InOmega}
\end{equation}
which predicts higher-order harmonics generation for \hbox{$\ell>1$}.

In the following we give the equivalent expressions of
\cref{eqn:Appendix:NLRTR:Hamiltonian,eqn:Appendix:NLRTR:vonNeumann,eqn:Appendix:NLRTR:rho,eqn:Appendix:NLRTR:CommutatorIdentity,eqn:Appendix:NLRTR:A,eqn:Appendix:NLRTR:GeneralizedKuboSusceptibility}
in imaginary times and frequencies.

\subsection{NLRT in imaginary time and frequency}
\label{ssec:imaginary-nlrt}

Similar derivations are found in standard text books like
\cite[Ch.~7.2]{AltlandSimons2010}. In fact, we follow~\cite{AltlandSimons2010} for the
most part, but keep the terms of higher order.

In imaginary time the Hamiltonian is given as
\begin{equation}
    \HHH(\tau) = \HH - \sum_j \AA_j F_j(\tau).
    \label{eqn:Appendix:NLRTI:Hamiltonian}
\end{equation}
The analogous action is
\begin{equation}
    \calS[\cgdag, \cg, \vect F] = S[\cgdag, \cg] - \sum_{j a a'} \int_{0}^{\beta} \!\!\! \dd \tau \, F_j(\tau) \cgdag_a(\tau) A_j^{aa'} \cg_{a'}(\tau)
    \label{eqn:Appendix:NLRTI:Action}
\end{equation}
where $\cgdag(\tau)$ and $\cg(\tau)$ are Grassmann-valued fields corresponding to the
eigenvalues of $\cdag(\tau)$ and $\cc(\tau)$ with respect to coherent states. $A_j^{aa'}$
are matrix elements of $\AA_j$: $\AA_j = \sum_{aa'} \cdag_a A_j^{aa'} \cc_{a'}$. The
derivation does not hinge $\AA_j$ being a one-particle operator, one can equally well
assume $\AA_j$ to be an arbitrary (hermitian) $n$-particle operator. The expectation value
$\ev*{\AA_i(\tau)}$ is best expressed in the path-integral formalism:
\begin{align}
    \ev*{ \dots}_{\vect{F}} &= \calZ^{-1} \int \calD \cgdag \calD \cg \ee^{-\calS[\cgdag, \cg, \vect{F}]} \dots
    \label{eqn:Appendix:NLRTI:PathintegralBasics01}
    \\
    \calZ &= \int \calD \cgdag \calD \cg \ee^{-\calS[\cgdag, \cg, \vect{F}]}
    \label{eqn:Appendix:NLRTI:PathintegralBasics02}
    \\
    \ev*{\AA_i(\tau)}_{\vect F} &= \fdv{F_i(\tau)} \log \calZ
    \label{eqn:Appendix:NLRTI:PathintegralBasics03}
\end{align}
This allows us to express $\ev*{\AA_i(\tau)}_{\vect F}$ as a functional Taylor series:
\begin{equation}
    \begin{array}{rl}
        \ev*{\AA_i(\tau)}_{\vect F} =& \sum_{\ell=0}^{\infty} \ev*{\Delta \AA_i(\tau)}^{({\ell})}_{\vect F}\\
        \ev*{\Delta \AA_i(\tau)}^{({\ell})}_{\vect F} =& \frac{1}{\ell!}
        \int_{0}^{\beta} \dd \tau_1 \int_{0}^{\beta} \dd \tau_2 \dots \int_{0}^{\beta} \dd \tau_\ell \\
        & \sum_{j_1, j_2, \dots, j_\ell} \\
        &F_{j_1}(\tau_1) F_{j_2}(\tau_2) \dots F_{j_\ell}(\tau_\ell) \\
        & \chicon_{j_1 j_2 \dots j_\ell i}(\tau_1, \tau_2, \dots, \tau_\ell, \tau),
    \end{array}
    \label{eqn:Appendix:NLRTI:ExpansionOfAinF}
\end{equation}
where the ($\ell+1$)-point susceptibility reads:
\begin{multline}
    \chicon_{j_1 j_2 \dots j_\ell i}(\tau_1, \tau_2, \dots, \tau_\ell, \tau) \\
    = \fdv{F_{j_1}(\tau_1)} \fdv{F_{j_2}(\tau_2)} \dots \fdv{F_{j_\ell}(\tau_\ell)}
    \ev*{\AA_i(\tau)}_{\vect F} \eval_{\vect F=0}
    \label{eqn:Appendix:NLRTI:ConnectedSusceptibilityOrderL}
\end{multline}
All functional derivatives can be exchanged with one another. Calculating a susceptibility
of order $\ell$ is a simple exercise in differentiation. The first-order term is the usual
linear response function
\begin{equation}
    \begin{array}{rcl}
        \chicon_{i j}(\tau, \tau') &=&
        \fdv{F_{i}(\tau)} \fdv{F_{j}(\tau')} \log \calZ \eval_{\vect F=0} \\
        &=& -\calZ^{-2} [ \fdv{F_i(\tau)} \calZ] [ \fdv{F_j(\tau')} \calZ] \eval_{\vect F=0} \\
        & & + \calZ^{-1} [ \fdv{F_i(\tau)} \fdv{F_j(\tau')} \calZ] \eval_{\vect F=0} \\
        &=& - \ev*{\AA_i} \ev*{\AA_j} + \ev*{\TT \AA_i(\tau) \AA_j(\tau')} \; .
    \end{array}
    \label{eqn:Appendix:NLRTI:ConnectedSusceptibilityOrder1}
\end{equation}
The second-order term reads:
\begin{equation}
    \begin{array}{rcl}
        \chicon_{i j_1 j_2}(\tau, \tau_1, \tau_2) &=&
        \fdv{F_{i}(\tau)} \fdv{F_{j_1}(\tau_1)} \fdv{F_{j_2}(\tau_2)} \log \calZ \eval_{\vect F=0} \\
        &\hspace{-11em}=&\hspace{-5.50em} 2 \calZ^{-3} [ \fdv{F_i(\tau)} \calZ] [ \fdv{F_{j_1}(\tau_1)} \calZ] [ \fdv{F_{j_2}(\tau_2)} \calZ]\eval_{\vect F=0} \\
        & &\hspace{-5.50em} - \calZ^{-2} [ \fdv{F_{i}(\tau)} \calZ] [ \fdv{F_{j_1}(\tau_1)} \fdv{F_{j_2}(\tau_2)} \calZ] \eval_{\vect F=0} \\
        & &\hspace{-5.50em} - \calZ^{-2} [ \fdv{F_{j_1}(\tau_1)} \calZ] [ \fdv{F_{i}(\tau)} \fdv{F_{j_2}(\tau_2)} \calZ] \eval_{\vect F=0} \\
        & &\hspace{-5.50em} - \calZ^{-2} [ \fdv{F_{j_2}(\tau_2)} \calZ] [ \fdv{F_{i}(\tau)} \fdv{F_{j_1}(\tau_1)} \calZ] \eval_{\vect F=0} \\
        & &\hspace{-5.50em} + \calZ^{-1} [ \fdv{F_i(\tau)} \fdv{F_{j_1}(\tau_1)} \fdv{F_{j_2}(\tau_2)} \calZ] \eval_{\vect F=0} \\
        &\hspace{-11em}=&\hspace{-5.50em} +2\ev*{\AA_i} \ev*{\AA_{j_1}} \ev*{\AA_{j_2}} \\
        & &\hspace{-5.50em} - \ev*{\AA_i } \ev*{\TT \AA_{j_1}(\tau_1) \AA_{j_2}(\tau_2)} \\
        & &\hspace{-5.50em} - \ev*{\AA_{j_1}} \ev*{\TT \AA_{i}(\tau) \AA_{j_2}(\tau_2)} \\
        & &\hspace{-5.50em} - \ev*{\AA_{j_2}} \ev*{\TT \AA_{i}(\tau) \AA_{j_1}(\tau_1)} \\
        & &\hspace{-5.50em} + \ev*{\TT \AA_{i}(\tau) \AA_{j_1}(\tau_1) \AA_{j_2}(\tau_2)},
    \end{array}
    \label{eqn:Appendix:NLRTI:ConnectedSusceptibilityOrder2}
\end{equation}
which after some reordering and plugging in
\cref{eqn:Appendix:NLRTI:ConnectedSusceptibilityOrder1} can be written as
\begin{multline}
    \chicon_{i j_1 j_2}(\tau, \tau_1, \tau_2) \\
    \begin{aligned}
        = {} & \ev*{\TT \AA_i(\tau) \AA_{j_1}(\tau_1) \AA_{j_2}(\tau_2)} \\
        & {} - \ev*{\AA_i} \chicon_{j_1 j_2}(\tau_1, \tau_2)
            - \ev*{\AA_{j_1}} \chicon_{i j_2}(\tau, \tau_2) \\
        & {} - \ev*{\AA_{j_2}} \chicon_{i j_1}(\tau, \tau_1)
            - \ev*{\AA_i} \ev*{\AA_{j_1}} \ev*{\AA_{j_2}}.
    \end{aligned}
    \label{eqn:Appendix:NLRTI:ConnectedSusceptibilityOrder2-2}
\end{multline}
\Cref{eqn:Appendix:NLRTI:ConnectedSusceptibilityOrder1,eqn:Appendix:NLRTI:ConnectedSusceptibilityOrder2,eqn:Appendix:NLRTI:ConnectedSusceptibilityOrder2-2}
conclude the derivation of \cref{eq:chi,eq:chi-2} in the main text.
     \section{Implementation details for three-particle correlators and second-order response
         functions}
\label{sec:3p-details}

In this section we present all the details and explicit equations, purposefully omitted
for brevity in \cref{sec:theory}, to get from two- and three-particle correlators to the
second-order density, density-magnetic, and chiral response functions. This is interesting
if one actually wants to implement the computations since, in our case,
\program{w2dynamics} can only directly measure the spin correlators.

We start by giving explicit formulas for \cref{eq:3p-corr}
\begin{align}
    \begin{split}
        \chinnn = {}
            & 2 \left(\chi_{\up\up\up} + \chi_{\up\up\down}
                      + \chi_{\up\down\up} + \chi_{\down\up\up}\right),
    \end{split}
    \label{eq:chi-nnn-spin}
    \\
    \begin{split}
        \chinzz = {}
            & 2 \left(\chi_{\up\up\up} - \chi_{\up\up\down}
                      - \chi_{\up\down\up} + \chi_{\down\up\up}\right) \\
        = {}
            & 2 \left(\chi_{\up\wbar{\up\down}} + \chi_{\down\wbar{\up\down}}\right),
    \end{split}
    \label{eq:chi-nzz-spin}
    \\
    \begin{split}
        \chixyz = {}
            & 2 \left(\chi_{\wbar{\up\down}\up} - \chi_{\wbar{\up\down}\down} \right),
    \end{split}
    \label{eq:chi-xyz-spin}
\end{align}
where we use SU(2) symmetry, introduce the second-order spin susceptibilities
\begin{equation}
    \chi_{\sigma_1 \dots \sigma_6} = \conn \corr_{\sigma_1 \dots \sigma_6},
\end{equation}
and generalize the compact spin notation from the two- to the three-particle level:
\begin{align}
    \sigma_1 \sigma_2 \sigma_3
        & = \sigma_1 \sigma_1 \sigma_2 \sigma_2 \sigma_3 \sigma_3,
    \label{eq:123}
    \\
    \sigma_1 \wbar{\sigma_2 \sigma_3}
        & = \sigma_1 \sigma_1 \sigma_2 \sigma_3 \sigma_3 \sigma_2,
    \label{eq:123-bar}
    \\
    \wbar{\sigma_1} \sigma_2 \wbar{\sigma_3}
        & = \sigma_1 \sigma_3 \sigma_2 \sigma_2 \sigma_3 \sigma_1,
    \label{eq:1-bar-23-bar}
    \\
    \wbar{\sigma_1 \sigma_2} \sigma_3
        & = \sigma_1 \sigma_2 \sigma_2 \sigma_1 \sigma_3 \sigma_3.
    \label{eq:12-bar-3}
\end{align}
Since subtracting the disconnected parts of the full correlators is a linear operation
\cref{eq:chi-nnn-spin,eq:chi-nzz-spin,eq:chi-xyz-spin} also hold when replacing $\chi$
with $\corr$.

\Cref{sec:symmetries-of-3p-corr} shows that the 20 non-vanishing spin components of the
three-particle quantities can be reduced to just three independent ones, namely
$\up\up\up$, $\up\up\down$ and $\up\!\wbar{\up\down}$. All other components can be
calculated from these by applying SU(2), swapping or time-reversal symmetry. This makes
numerical computations much cheaper. Exploiting this, we rewrite
\cref{eq:chi-nnn-spin,eq:chi-nzz-spin,eq:chi-xyz-spin} in Matsubara space as
\begin{align}
    \begin{split}
        \chinnn^{\omega_1 \omega_2} = {}
        & 2 \left(\chi^{\omega_1 \omega_2}_{\up\up\up}
                  + \chi^{\omega_1 \omega_2}_{\up\up\down}
                  + \chi^{\omega_1 \omega_3}_{\up\up\down}
                  + \chi^{\omega_3 \omega_2}_{\up\up\down}\right),
    \end{split}
    \label{eq:chi-nnn-spin-2}
    \\
    \begin{split}
        \chinzz^{\omega_1 \omega_2} = {}
            & 2 \left(\chi^{\omega_1 \omega_2}_{\up\up\up}
                      - \chi^{\omega_1 \omega_2}_{\up\up\down}
                      - \chi^{\omega_1 \omega_3}_{\up\up\down}
                      + \chi^{\omega_3 \omega_2}_{\up\up\down}\right) \\
        = {}
            & 2 \left(\chi^{\omega_1 \omega_2}_{\up\wbar{\up\down}}
                      + \chi^{\omega_1 \omega_3}_{\up\wbar{\up\down}}\right),
    \end{split}
    \label{eq:chi-nzz-spin-2}
    \\
    \begin{split}
        \chixyz^{\omega_1 \omega_2} = {}
        & 2 \left(\chi^{\omega_3 \omega_1}_{\up\wbar{\up\down}}
                  - \chi^{\omega_3 \omega_2}_{\up\wbar{\up\down}}\right),
    \end{split}
    \label{eq:chi-xyz-spin-2}
\end{align}
where $\omega_3 = -\omega_1 - \omega_2$.

For the second-order spin susceptibilities the explicit form of \cref{eq:chi-2-omega}
reads
\begin{align}
    \begin{split}
        \chi^{\omega_1 \omega_2}_{\sigma_1 \sigma_2 \sigma_3} = {}
            & \corr_{\sigma_1 \sigma_2 \sigma_3}^{\omega_1 \omega_2}
                - \delta_{\omega_1 0} \delta_{\omega_2 0} \beta^2
                \ev{\nn_{\sigma_1}} \ev{\nn_{\sigma_2}} \ev{\nn_{\sigma_3}} \\
            & {} - \delta_{\omega_2 0} \beta \ev{\nn_{\sigma_2}}
                \chi_{\sigma_3 \sigma_1}^{\omega_3} \\
            & {} - \delta_{\omega_3 0} \beta \ev{\nn_{\sigma_3}}
                \chi_{\sigma_1 \sigma_2}^{\omega_1} \\
            & {} - \delta_{\omega_1 0} \beta \ev{\nn_{\sigma_1}}
                \chi_{\sigma_2 \sigma_3}^{\omega_2},
    \end{split}
    \label{eq:chi-spin-123-omega} \\
    \begin{split}
        \chi^{\omega_1 \omega_2}_{\sigma_1 \wbar{\sigma_2 \sigma_3}} = {}
            & \corr_{\sigma_1 \wbar{\sigma_2 \sigma_3}}^{\omega_1 \omega_2}
                - \delta_{\omega_1 0} \beta \ev{\nn_{\sigma_1}}
                \chi_{\wbar{\sigma_3 \sigma_2}}^{\omega_2},
    \end{split}
    \label{eq:chi-spin-123bar-omega} \\
    \begin{split}
        \chi^{\omega_1 \omega_2}_{\wbar{\sigma_1} \sigma_2 \wbar{\sigma_3}} = {}
            & \corr_{\wbar{\sigma_1} \sigma_2 \wbar{\sigma_3}}^{\omega_1 \omega_2}
                - \delta_{\omega_2 0} \beta \ev{\nn_{\sigma_2}}
                \chi_{\wbar{\sigma_1 \sigma_3}}^{\omega_3},
    \end{split}
    \label{eq:chi-spin-1bar23bar-omega} \\
    \begin{split}
        \chi^{\omega_1 \omega_2}_{\wbar{\sigma_1 \sigma_2} \sigma_3} = {}
            & \corr_{\wbar{\sigma_1 \sigma_2} \sigma_3}^{\omega_1 \omega_2}
                - \delta_{\omega_3 0} \beta \ev{\nn_{\sigma_3}}
                \chi_{\wbar{\sigma_2 \sigma_1}}^{\omega_1}.
    \end{split}
    \label{eq:chi-spin-12bar3-omega}
\end{align}
Similarly, using \cref{eq:chi-omega} yields
\begin{align}
    \chi^\omega_{\sigma_1 \sigma_2}
        & = \ev*{\TT \nn_{\sigma_1} \nn_{\sigma_2}}^\omega
            - \delta_{\omega 0} \beta \ev*{\nn_{\sigma_1}} \ev*{\nn_{\sigma_2}} \\
    \chi^\omega_{\wbar{\sigma_1 \sigma_2}}
        & = \ev*{\TT \cdag_{\sigma_1} \cc_{\sigma_2}
                 \cdag_{\sigma_2} \cc_{\sigma_1}}^\omega
\end{align}
for the linear spin susceptibilities. Together these equations complete the set of
explicit formulas necessary to compute the second-order density, density-magnetic, and
chiral response functions from two- and three-particle spin correlators.

Combining \cref{eq:chi-xyz-spin} and \cref{eq:chi-spin-12bar3-omega} shows that the
disconnected terms for $\chixyz$ cancel, which means that it is directly given by
$\corr_{xyz}$. This is similar to the two-particle case where the magnetic response
function $\chi_{zz}$ equals the full correlator $\corr_{zz} = \ev*{\TT \zz \zz}$ [see
\cref{eq:chi-zz}] because there are no disconnected terms either.

In the special case of half-filling, i.e., $\ev*{\nn_\sigma} = 1/2 = 1 - \ev*{\nn_\sigma}$
we can further compute
\begin{align}
    \corr_{\sigma_1 \sigma_2 \sigma_3} = {}
        & \ev*{\TT (1 - \nn_{\sigma_1}) (1 - \nn_{\sigma_2}) (1 - \nn_{\sigma_3})} \\
    \begin{split}
        = {}
            & \ev*{1 - \nn_{\sigma_1} - \nn_{\sigma_2} - \nn_{\sigma_3}} \\
            & + \ev*{\TT (\nn_{\sigma_1} \nn_{\sigma_2} + \nn_{\sigma_1} \nn_{\sigma_3}
                + \nn_{\sigma_2} \nn_{\sigma_3})} \\
            & - \ev*{\TT \nn_{\sigma_1} \nn_{\sigma_2} \nn_{\sigma_3}},
    \end{split} \\
    2 \corr_{\sigma_1 \sigma_2 \sigma_3} = {}
        & \sum_{i < j} \ev*{\TT \nn_{\sigma_i} \nn_{\sigma_j}} - \frac{1}{2} \, .
\end{align}
This shows that the full, density-like, three-particle spin correlators only consist of
disconnected terms for half-filling, or equivalently $\chi_{\sigma_1 \sigma_2 \sigma_3}$
vanishes. Looking at \cref{eq:chi-nnn-spin,eq:chi-nzz-spin} this also implies that the
second-order density and density-magnetic response functions vanish at half-filling.
     \section{Frequency notations of three-particle diagrams}
\label{sec:frequency-notations-of-3p-diagrams}

\begin{figure}
    \centering
    \includeinkscapefigure{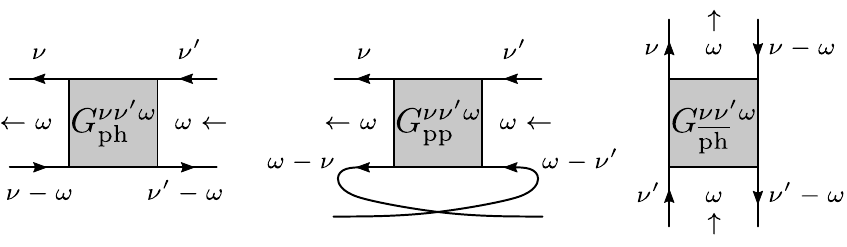}
    \caption{Frequency notations for the two-particle Green's function.}
    \label{fig:2p-frequency-notation}
\end{figure}

In the two-particle case there are three channels each with its own frequency notation. It
is chosen such that the in- and outgoing particle--particle (pp) or particle--hole (ph)
pairs have a total energy of $\omega$. The corresponding diagrams for the two-particle
Green's function are shown in \cref{fig:2p-frequency-notation}.

Making similar pairwise connections as in \cref{fig:2p-frequency-notation}, the number of
different frequency notations for an $n$-particle function is equal to the number of
different ways that the $2n$ points can be connected into pairs. This number of
possibilities $p$ is given by the double factorial
\begin{equation}
    p(n) = (2n - 1)!! = \prod_{k=1}^n (2k - 1),
    \label{eq:number-of-channels}
\end{equation}
because with $2k$ remaining points there are $2k - 1$ possibilities to connect an
arbitrarily chosen point to one of the $2k - 1$ other ones. The number of pure ph channels
is $n!$ since in this case each of the $n$ creation operators must be paired with one of
the $n$ annihilation operators and there are $n!$ unique ways to do that.

\begin{figure}
    \centering
    \includeinkscapefigure[width=\columnwidth]{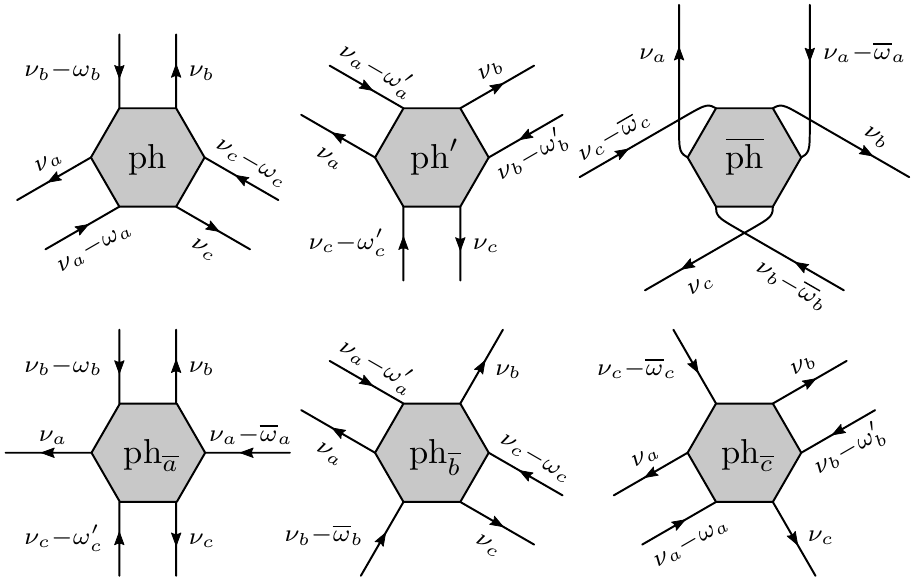}
    \caption{Diagrammatic representation of the six ph notations of three-particle
             diagrams.}
    \label{fig:ph-notation}
\end{figure}

\begin{figure}
    \centering
    \includeinkscapefigure[width=\columnwidth]{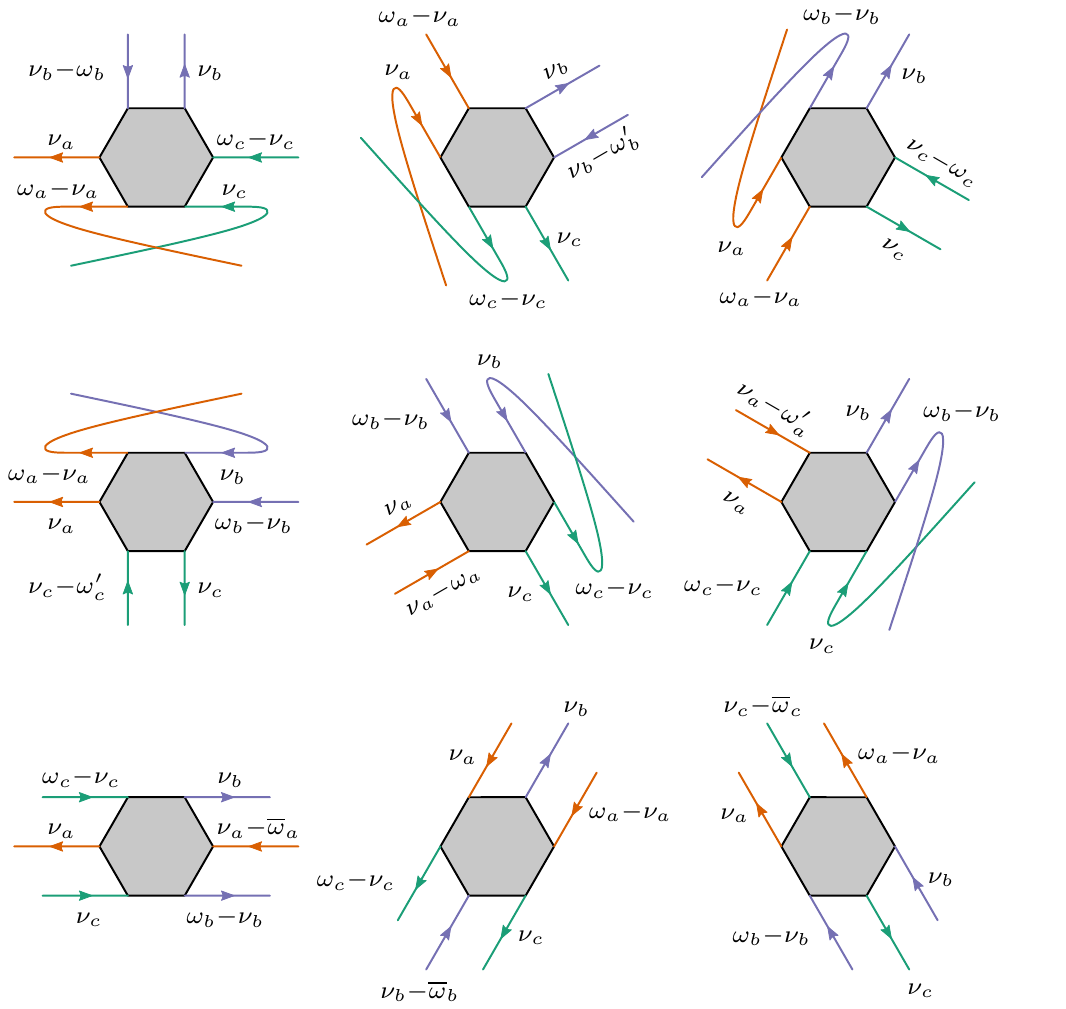}
    \caption{Diagrammatic representation of the nine pp notations of three-particle
             diagrams.}
    \label{fig:pp-notation}
\end{figure}

For the three-particle case this results in 15 different frequency notations which are
shown in \cref{tab:frequency-notations}.
\begin{table}
    \centering
    \setlength{\tabcolsep}{2pt}
    \begin{tabular}{@{}lcccccc@{}}
        \toprule
        channel & $\nu_{1}$ & $\nu_{2}$ & $\nu_{3}$ & $\nu_{4}$ & $\nu_{5}$ & $\nu_{6}$ \\
        \midrule
        ph & $\nu_a - \omega_a$ & $\nu_a$ & $\nu_b - \omega_b$ & $\nu_b$ & $\nu_c -
            \omega_c$ & $\nu_c$ \\
        $\t{ph}'$ & $\nu_c - \omega'_c$ & $\nu_a$ & $\nu_a - \omega'_a$ & $\nu_b$ & $\nu_b
                    - \omega'_b$ & $\nu_c$ \\
        $\wbar{\t{ph}}$ & $\nu_b - \wbar{\omega}_b$ & $\nu_a$ & $\nu_c - \wbar{\omega}_c$
                        & $\nu_b$ & $\nu_a - \wbar{\omega}_a$ & $\nu_c$ \\
        $\t{ph}_{\wbar{a}}$ & $\nu_c - \omega'_c$ & $\nu_a$ & $\nu_b - \omega_b$ & $\nu_b$
                            & $\nu_a - \wbar{\omega}_a$ & $\nu_c$ \\
        $\t{ph}_{\wbar{b}}$ & $\nu_b - \wbar{\omega}_b$ & $\nu_a$ & $\nu_a - \omega'_a$ &
                            $\nu_b$ & $\nu_c - \omega_c $ & $\nu_c$ \\
        $\t{ph}_{\wbar{c}}$ & $\nu_a - \omega_a$ & $\nu_a$ & $\nu_c - \wbar{\omega}_c$ &
                            $\nu_b$ & $\nu_b - \omega'_b$ & $\nu_c$ \\
        \midrule
        $\t{pp}_{24-13}$ & $\nu_c$ & $\nu_a$ & $\nu_b - \omega_b$ & $\nu_b$ & $\omega_c -
                            \nu_c$ & $\omega_a - \nu_a$ \\
        $\t{pp}_{26-13}$ & $\nu_a$ & $\omega_c - \nu_c$ & $\omega_a - \nu_a$ & $\nu_b$ &
                            $\nu_b - \omega'_b$ & $\nu_c$ \\
        $\t{pp}_{26-15}$ & $\omega_a - \nu_a$ & $\omega_b - \nu_b$ & $\nu_a$ & $\nu_b$ &
                            $\nu_c - \omega_c$ & $\nu_c$ \\
        $\t{pp}_{46-15}$ & $\nu_c - \omega'_c$ & $\nu_a$ & $\nu_b$ & $\omega_a - \nu_a$ &
                            $\omega_b - \nu_b$ & $\nu_c$ \\
        $\t{pp}_{46-35}$ & $\nu_a - \omega_a$ & $\nu_a$ & $\omega_b - \nu_b$ & $\omega_c -
                            \nu_c$ & $\nu_b$ & $\nu_c$ \\
        $\t{pp}_{24-35}$ & $\omega_c - \nu_c$ & $\nu_a $ & $\nu_a - \omega'_a$ & $\nu_b$ &
                            $\nu_c$ & $\omega_b - \nu_b$ \\
        \midrule
        $\t{pp}_{26-35}$ & $\nu_c$ & $\nu_a$ & $\omega_c - \nu_c$ & $\nu_b$ & $\nu_a -
                            \wbar{\omega}_a$ & $\omega_b - \nu_b$ \\
        $\t{pp}_{46-13}$ & $\nu_b - \wbar{\omega}_b$ & $\omega_c - \nu_c$ & $\nu_a$ &
                            $\nu_b$ & $\omega_a - \nu_a$ & $\nu_c$ \\
        $\t{pp}_{24-15}$ & $\omega_b - \nu_b$ & $\nu_a$ & $\nu_c - \wbar{\omega}_c$ &
                            $\omega_a - \nu_a$ & $\nu_b$ & $\nu_c$ \\
        \bottomrule
    \end{tabular}
    \caption{The 15 different frequency notations of three-particle diagrams.}
    \label{tab:frequency-notations}
\end{table}
They can be divided into six ph channels and nine pp channels. A diagrammatic
representation is given in \cref{fig:ph-notation,fig:pp-notation}, depicting the ph and pp
channels respectively.
     \section{Decomposition of the three-particle Green's function}
\label{sec:decomposition-of-g3}

\begin{figure*}
    \centering
    \includeinkscapefigure{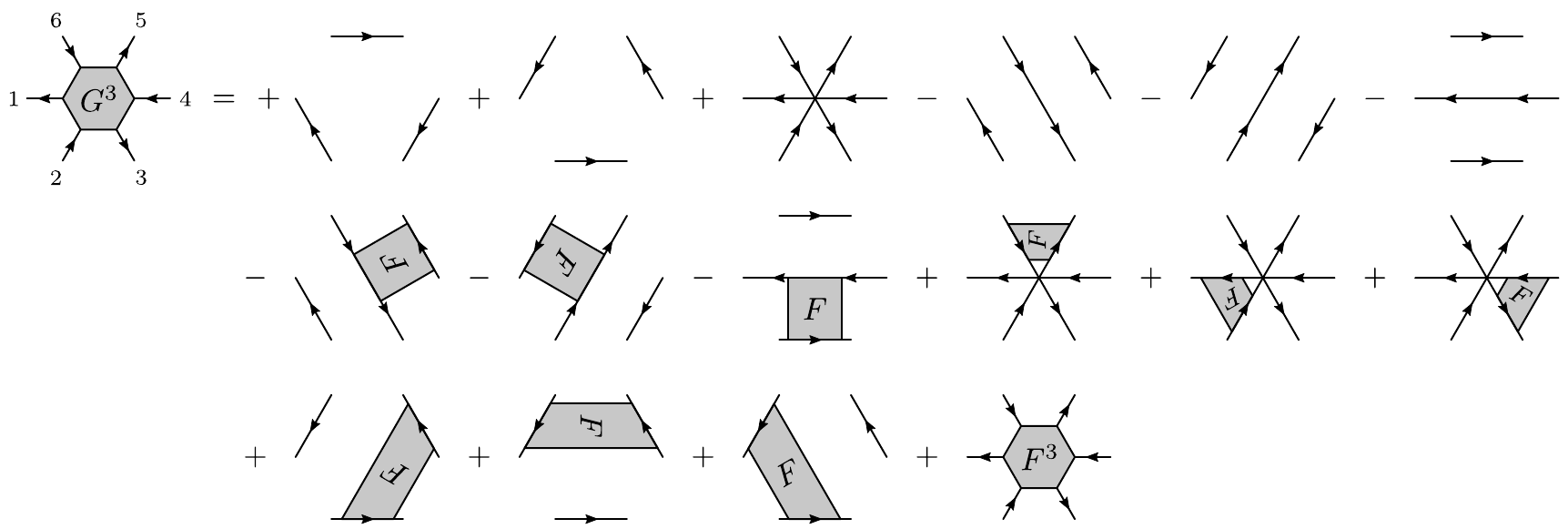}
    \caption{Diagrammatic representation of the decomposition of the three-particle
    Green's function as given in \cref{eq:g3-decomposition}. The last term introduces the
    full three-particle vertex $F^3$.}
    \label{fig:g3-decomposition}
\end{figure*}

\begin{figure}
    \centering
    \includeinkscapefigure{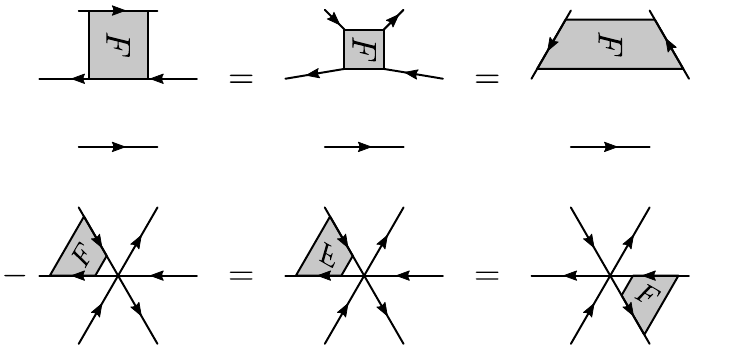}
    \caption{Some topologically equivalent diagrams that appear in the decomposition of
             the three-particle Green's function shown in \cref{fig:g3-decomposition}. In
             the second line we use the crossing symmetry of $F$ for the first identity.}
    \label{fig:equivalent-diagrams}
\end{figure}

The general idea is to decompose the expectation value of an $n$-particle Green's function
into all possible sets of connected tuples of creation and annihilation operators. If we
denote the connected tuples by underlined expectation values the one-particle case is
trivially written as
\begin{equation}
    G^1_{12} = -\ev*{\TT \cc_1 \cdag_2} = -\uev{\cc_1 \cdag_2},
    \label{eq:g1-decomposition}
\end{equation}
where we condensed all arguments and indices of each operator into a single numeric index.
On the two-particle level we get
\begin{equation}
    \begin{split}
        G^2_{1234} = {} & \ev*{\TT \cc_1 \cdag_2 \cc_3 \cdag_4} \\
        = {}
            & \uev{\cc_1 \cdag_2 \cc_3 \cdag_4} + \uev{\cc_1 \cdag_2}
                \uev{\cc_3 \cdag_4} \\
            & {} - \uev{\cc_1 \cdag_4} \uev{\cc_3 \cdag_2},
    \end{split}
    \label{eq:g2-decomposition}
\end{equation}
where the first term on the right-hand side contains the full two-particle vertex $F$. The
diagrammatic representation of this equation is given in \cref{fig:g2-decomposition} which
also shows that we choose the following convention for the sign of $F$:
\begin{figure}
    \centering
    \includeinkscapefigure{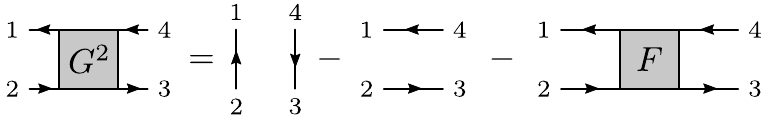}
    \caption{Diagrammatic representation of the decomposition of the two-particle Green's
             function $G^2$, as given in \cref{eq:g2-decomposition}. The last term
             introduces the full two-particle vertex $F$.}
    \label{fig:g2-decomposition}
\end{figure}
\begin{equation}
    \uev{\cc \cdag \cc \cdag} = - GG F GG.
    \label{eq:sign-of-f}
\end{equation}
So far this is nothing new for people who are well versed in diagrammatics. The
decomposition of the three-particle Green's function, however, is less well known and
simple which is why we explicitly present it in this section. Applying the same method as
before we get
\begin{widetext}
    \begin{equation}
        \begin{split}
            -\ev*{\TT \cc_1 \cdag_2 \cc_3 \cdag_4 \cc_5 \cdag_6} = {}
            &
            - \uev{\cc_1 \cdag_2} \uev{\cc_3 \cdag_4} \uev{\cc_5 \cdag_6}
            - \uev{\cc_1 \cdag_6} \uev{\cc_3 \cdag_2} \uev{\cc_5 \cdag_4}
            - \uev{\cc_1 \cdag_4} \uev{\cc_3 \cdag_6} \uev{\cc_5 \cdag_2}
            \\
            &
            + \uev{\cc_1 \cdag_2} \uev{\cc_3 \cdag_6} \uev{\cc_5 \cdag_4}
            + \uev{\cc_1 \cdag_6} \uev{\cc_3 \cdag_4} \uev{\cc_5 \cdag_2}
            + \uev{\cc_1 \cdag_4} \uev{\cc_3 \cdag_2} \uev{\cc_5 \cdag_6}
            \\
            &
            - \uev{\cc_1 \cdag_2} \uev{\cc_3 \cdag_4 \cc_5 \cdag_6}
            - \uev{\cc_3 \cdag_4} \uev{\cc_5 \cdag_6 \cc_1 \cdag_2}
            - \uev{\cc_5 \cdag_6} \uev{\cc_1 \cdag_2 \cc_3 \cdag_4}
            \\
            &
            + \uev{\cc_1 \cdag_4} \uev{\cc_3 \cdag_2 \cc_5 \cdag_6}
            + \uev{\cc_3 \cdag_6} \uev{\cc_5 \cdag_4 \cc_1 \cdag_2}
            + \uev{\cc_5 \cdag_2} \uev{\cc_1 \cdag_6 \cc_3 \cdag_4}
            \\
            &
            + \uev{\cc_1 \cdag_6} \uev{\cc_3 \cdag_4 \cc_5 \cdag_2}
            + \uev{\cc_3 \cdag_2} \uev{\cc_5 \cdag_6 \cc_1 \cdag_4}
            + \uev{\cc_5 \cdag_4} \uev{\cc_1 \cdag_2 \cc_3 \cdag_6}
            \\
            & -\uev{\cc_1 \cdag_2 \cc_3 \cdag_4 \cc_5 \cdag_6}
            .
        \end{split}
        \label{eq:g3-decomposition}
    \end{equation}
\end{widetext}
The corresponding diagrams are depicted in \cref{fig:g3-decomposition} where we introduce
the full three-particle vertex $F^3$.
Its sign is chosen as
\begin{equation}
    -\uev{\cc_1 \cdag_2 \cc_3 \cdag_4 \cc_5 \cdag_6} = GGG F^3 GGG.
    \label{eq:sign-of-f3}
\end{equation}
Looking at \cref{fig:g3-decomposition} it might seem that some diagrams are
\enquote{obviously} missing, but they are in fact just topologically equivalent to some of
the already present ones. Examples for this are shown in \cref{fig:equivalent-diagrams}.
     \section{Asymptotic behavior of \texorpdfstring{$\chitwo$}{chi2}}
\label{sec:asymptotic-behavior-of-chi-2}

Using the Lehmann representation one can show that bosonic, two-particle correlation
functions can be expanded in the following series:
\begin{equation}
    \begin{split}
        \ev*{\AA_i \AA_j}^z = {} & -\frac{1}{z} \ev*{\comm*{\AA_i}{\AA_j}} \\
        & + \frac{1}{z^2} \ev*{\comm*{\comm*{\AA_i}{\HH}}{\AA_j}} + \dots,
    \end{split}
    \label{eq:z-expansion}
\end{equation}
where $z$ is a complex frequency, $\HH$ is the Hamiltonian and $\comm*{\cdot}{\cdot}$
denotes the commutator (see also \cite[Appendix C]{Krien2017}). Since
\begin{align}
    \chinnn^{0 \omega}
    & = -\pdv{\epsilon} \chi_{nn}^\omega = \pdv{\epsilon} \pqty{
        \corr_{nn}^\omega - \delta_{\omega 0} \beta \ev*{\nn}^2} \\
    & = \pdv{\epsilon} \corr_{nn}^\omega,
    \label{eq:chi-nnn-0-omega}
    \\
    \chinzz^{0 \omega} & = \pdv{\epsilon} \chi_{zz}^{\omega}
        = \pdv{\epsilon} \corr_{zz}^{\omega},
    \label{eq:chi-nzz-0-omega}
    \\
    \chixyz^{0 \omega} & = \pdv{h_x} \chi_{yz}^{\omega} = \pdv{h_x} \corr_{yz}^{\omega},
    \label{eq:chi-xyz-0-omega}
\end{align}
with Matsubara frequencies $\omega$, evaluating the expansion at $z = \ii \omega$ can be
used to obtain the asymptotic behavior of slices of the second-order response functions:
\begin{align}
    \chinnn^{0 \omega} \approx {}
        & {} -\frac{1}{(\ii \omega)^2} \pdv{\epsilon} \ev*{\comm*{\comm*{\nn}{\HH}}{\nn}},
    \label{eq:chi-nnn-asymptote-1}
    \\
    \chinzz^{0 \omega} \approx {}
        & {} -\frac{1}{(\ii \omega)^2} \pdv{\epsilon} \ev*{\comm*{\comm*{\zz}{\HH}}{\zz}},
    \label{eq:chi-nzz-asymptote-1}
    \\
    \chixyz^{0 \omega} \approx {}
        & {} -\frac{1}{\ii \omega} \pdv{h_x} \ev*{\comm*{\yy}{\zz}}.
    \label{eq:chi-xyz-asymptote-1}
\end{align}
The density and density-magnetic channels do not have a $1 / (\ii \omega)$ term since
$\comm{\nn}{\nn}$ and $\comm{\zz}{\zz}$ vanish. According to~\cite{Krien2017}, for an AIM
the commutators in \cref{eq:chi-nnn-asymptote-1,eq:chi-nzz-asymptote-1} are given by
\begin{equation}
    \begin{split}
        \ev*{\comm*{\comm*{\nn}{\HH_\text{AIM}}}{\nn}}
            & = \ev*{\comm*{\comm*{\zz}{\HH_\text{AIM}}}{\zz}} \\
        & = -\ev*{\HH_V}
            = -\frac{2}{\beta} \sum_{\sigma \nu} \Delta_\sigma^\nu G_\sigma^\nu
    \end{split}
\end{equation}
where $\HH_V$ is the hybridization term in the Hamiltonian of the AIM [last term in
\cref{eq:aim-hamiltonian}], $\Delta_\sigma^\nu$ is the hybridization function and
$G_\sigma^\nu$ is the one-particle Green's function of the impurity. Differentiating the
latter with respect to $\epsilon$ yields
\begin{align}
    \pdv{\epsilon} G_\sigma^\nu = {}
        & -\pdv{\epsilon} \ev*{\TT \cc_{\sigma}(\tau) \cdag_{\sigma}}^\nu \\
    = {}
        & {} \beta \ev*{\nn} G_\sigma^\nu \\
    & {} + \sum_{\sigma'} \ev*{
        \TT \nn_{\sigma'}(\tau') \cc_{\sigma}(\tau) \cdag_{\sigma}}^{\nu 0} \\
    = {}
        & {} \beta \ev*{\nn} G_\sigma^\nu \\
    & {} + \sum_{\sigma'} \ev*{
        \TT (1 - \cc_{\sigma'}(\tau') \cdag_{\sigma'}(\tau'))
        \cc_{\sigma}(\tau) \cdag_{\sigma}}^{\nu 0} \\
    = {}
        & {} \beta (\! \ev*{\nn} - 2) G_\sigma^\nu
        - \sum_{\sigma'} \Pthree_{\sigma' \sigma}^{\nu 0}
\end{align}
where $\Pthree$ is the partially contracted two-particle Green's function
\begin{equation}
    \begin{split}
        \Pthree^{\nu' \omega}
            & = \int_0^\beta \int_0^\beta G(\tau, \tau, \tau')
            \ee^{\ii (\omega \tau + (\nu' - \omega) \tau')} \dd{\tau} \dd{\tau'} \\
        & = \frac{1}{\beta} \sum_\nu G^{\nu \nu' \omega}.
    \end{split}
\end{equation}
With
\begin{equation}
    \pdv{h_x} \ev*{\comm{\yy}{\zz}} = 2 \ii \pdv{h_x} \ev*{\xx} = 2 \ii \chi_m^0
\end{equation}
we can finally write
\begin{align}
    \chinnn^{0 \omega} \approx {} & {} -\frac{1}{\omega^2} \pdv{\epsilon} \ev*{H_V}
    \label{eq:chi-nnn-asymptote-2}
    \\
    \chinzz^{0 \omega} \approx {} & {} -\frac{1}{\omega^2} \pdv{\epsilon} \ev*{H_V}
    \label{eq:chi-nzz-asymptote-2}
    \\
    \chixyz^{0 \omega} \approx {} & {} -\frac{2}{\omega} \chi_m^0,
    \label{eq:chi-xyz-asymptote-2}
\end{align}
where
\begin{equation}
    -\pdv{\epsilon} \ev*{H_V}
    = \frac{4}{\beta} \sum_\nu \Delta_\up^\nu \pqty{
        \Pthree_{\up\up}^{\nu 0} + \Pthree_{\up\down}^{\nu 0}
        + \beta (2 - \! \ev*{\nn}) G_\up^\nu},
    \label{eq:dhv-depsilon-1}
\end{equation}
and we use SU(2) symmetry.
     \section{Lehmann formula for the three-particle correlator}
\label{sec:lehmann}

In the atomic limit, $\HH_\text{AL} = \epsilon (\nn_\up + \nn_\down) - h (\nn_\up -
\nn_\down) + U \nn_\up \nn_\down$, where $\epsilon=-U/2$, we use the following Lehmann
formula for the three-particle correlation function:
\begin{equation}
    \begin{split}
        \corr(\tau_1, \tau_2) = {} &
            \ev{\TT \hat{\rho}_1(\tau_1) \hat{\rho}_2(\tau_2) \hat{\rho}_3(0)} \\
        = {} & \theta(\tau_1 - \tau_2) \sum_{i,j,k} w_i
            \ee^{\tau_1 E_{ij} + \tau_2 E_{jk}} \rho_1^{ij} \rho_2^{jk} \rho_3^{ki} \\
        & {} + \theta(\tau_2 - \tau_1) \sum_{i,j,k} w_i
            \ee^{\tau_2 E_{ij} + \tau_1 E_{jk}} \rho_2^{ij} \rho_1^{jk} \rho_3^{ki}.
    \end{split}
    \label{eq:x3lehmann_tau}
\end{equation}
Here, $\hat{\rho}_{1,2,3}$ are bosonic operators, $\theta$ is the Heaviside step function,
$E_{ij} = E_i - E_j$, $w_i = \ee^{-\beta E_i} / \mathcal{Z}$, $\mathcal{Z} = \sum_i
\ee^{-\beta E_i}$, and $\rho^{ij} = \matrixel{i}{\hat{\rho}}{j}$. The eigenstates of
$\HH_\text{AL}$ are $\ket{0}$, $\ket{\downarrow}$, $\ket{\uparrow}$, and
$\ket{\updownarrow}$ with eigenenergies $E_0=0$, $E_\down = \epsilon + h$, $E_\up =
\epsilon - h$, and $E_\updownarrow = U + 2 \epsilon$.

We transform $\corr(\tau_1, \tau_2)$ via \cref{eq:chiomom} to frequencies, taking care of
degeneracies:
\begin{equation}
    \corr^{\omega_1 \omega_2} =
        \mathcal{\corr}(\omega_1, \omega_2, \hat{\rho}_1, \hat{\rho}_2)
        + \mathcal{\corr}(\omega_2, \omega_1, \hat{\rho}_2, \hat{\rho}_1),
    \label{eq:x3lehmann_partial}
\end{equation}
\begin{widetext}
    \begin{equation}
        \begin{split}
            \mathcal{\corr}(\omega_x, \omega_y, \hat{\rho}_x, \hat{\rho}_y) = {} &
                \sum_{i,j,k} w_i \rho_x^{ij} \rho_y^{jk} \rho_3^{ki}
                \left\{\frac{1 - \delta(\ii \omega_y + E_{jk})}{\ii \omega_y + E_{jk}}
                       \left[-\frac{\ee^{\beta E_{ij}} - 1}{\ii \omega_x + E_{ij}}
                             \left(1 - \delta(\ii \omega_x + E_{ij})\right)
                             - \beta \delta(\ii \omega_x + E_{ij})\right.\right. \\
            & \left.
                + \frac{\ee^{\beta E_{ik}} - 1}{\ii \omega_x + \ii \omega_y + E_{ik}}
                \left(1 - \delta(\ii \omega_x + \ii \omega_y + E_{ik})\right)
                + \beta \delta(\ii \omega_x + \ii \omega_y + E_{ik})\right] \\
            & \left.
                + \left[\frac{\beta \ee^{\beta E_{ij}}}{\ii \omega_x + E_{ij}}
                        -\frac{\ee^{\beta E_{ij}} - 1}{(\ii \omega_x + E_{ij})^2}\right]
                \delta(\ii \omega_y + E_{jk})(1 - \delta(\ii \omega_x + E_{ij}))
                + \frac{\beta^2}{2} \delta(\ii \omega_y + E_{jk})
                    \delta(i \omega_x + E_{ij})\right\}.
        \end{split}
        \label{eq:x3lehmann}
    \end{equation}
\end{widetext}
Note that \cref{eq:x3lehmann_tau,eq:x3lehmann_partial,eq:x3lehmann} are not restricted to
the atomic limit.
     \section{Symmetries of the three-particle spin correlator}
\label{sec:symmetries-of-3p-corr}

There are ${6 \choose 3} = 20$ non-vanishing spin combinations for the full,
three-particle correlator. With the compact spin notation introduced in
\cref{eq:123,eq:123-bar,eq:1-bar-23-bar,eq:12-bar-3} they read
\begin{equation}
    \begin{aligned}
        & &
        & \up\up\up, &
        & \up\up\down, &
        & \up\down\up, &
        & \down\up\up, &
        \\
        & \pbs \wbar{\up} \nbs \up \nbs \wbar{\down}, &
        & \up \nbs \wbar{\up\down}, &
        & \up \nbs \wbar{\down\up}, &
        & \pbs \wbar{\up\down} \nbs \up, &
        & \pbs \wbar{\down\up} \nbs \up, &
        & \pbs \wbar{\down} \nbs \up \nbs \wbar{\up}, &
        \\
        & \pbs \wbar{\down} \nbs \down \nbs \wbar{\up}, &
        & \down \nbs \wbar{\down\up}, &
        & \down \nbs \wbar{\up\down}, &
        & \pbs \wbar{\down\up} \nbs \down, &
        & \pbs \wbar{\up\down} \nbs \down, &
        & \pbs \wbar{\up} \nbs \down \nbs \wbar{\down}, &
        \\
        & &
        & \down\down\down, &
        & \down\down\up, &
        & \down\up\down, &
        & \up\down\down. &
    \end{aligned}
\end{equation}
By using SU(2), swapping (SW) and time reversal (TR) symmetry
\begin{align}
    \corr^{\omega_1 \omega_2}_{\sigma_1 \dots \sigma_6} \overset{\text{SU(2)}}&{=}
    \corr^{\omega_1 \omega_2}_{-\sigma_1 \dots -\sigma_6},
    \\
    \corr^{\omega_1 \omega_2}_{\sigma_1 \dots \sigma_6} \overset{\text{SW12}}&{=}
    \corr^{\omega_2 \omega_1}_{\sigma_3 \sigma_4 \sigma_1 \sigma_2 \sigma_5 \sigma_6},
    \\
    \corr^{\omega_1 \omega_2}_{\sigma_1 \dots \sigma_6} \overset{\text{SW13}}&{=}
    \corr^{\omega_3 \omega_1}_{\sigma_5 \sigma_6 \sigma_3 \sigma_4 \sigma_1 \sigma_2},
    \\
    \corr^{\omega_1 \omega_2}_{\sigma_1 \dots \sigma_6} \overset{\text{SW23}}&{=}
    \corr^{\omega_1 \omega_3}_{\sigma_1 \sigma_2 \sigma_5 \sigma_6 \sigma_3 \sigma_4},
    \\
    \corr^{\omega_1 \omega_2}_{\sigma_1 \dots \sigma_6} \overset{\text{TR}}&{=}
    \corr^{\omega_3 \omega_2}_{\sigma_6 \dots \sigma_1},
\end{align}
where $-\up \, = \, \down$, $-\down \, = \, \up$, and $\omega_3 = -\omega_1 - \omega_2$,
they can be mapped to only three spin components, namely, $\up\up\up$, $\up\up\down$ and
$\up \nbs \wbar{\up\down}$. For seven of the first ten spin components the necessary
transformations look like
\begin{align}
    \corr_{\up\down\up}^{\omega_1 \omega_2}
    \overset{\text{SW23}}&{=} \corr_{\up\up\down}^{\omega_1 \omega_3}
    \\
    \corr_{\down\up\up}^{\omega_1 \omega_2}
    \overset{\text{SW13}}&{=} \corr_{\up\up\down}^{\omega_3 \omega_2}
    \\
    \corr_{\wbar{\up}\up\wbar{\down}}^{\omega_1 \omega_2}
    \overset{\text{SW12}}&{=}
    \corr_{\up\wbar{\up\down}}^{\omega_2 \omega_1}
    \\
    \corr_{\up\wbar{\down\up}}^{\omega_1 \omega_2}
    \overset{\text{SW23}}&{=} \corr_{\up\wbar{\up\down}}^{\omega_1 \omega_3}
    \\
    \corr_{\wbar{\up\down}\up}^{\omega_1 \omega_2}
    \overset{\text{SW23}}&{=} \corr_{\wbar{\up}\up\wbar{\down}}^{\omega_1 \omega_3}
    \overset{\text{SW12}}{=} \corr_{\up\wbar{\up\down}}^{\omega_3 \omega_1}
    \\
    \corr_{\wbar{\down\up}\up}^{\omega_1 \omega_2}
    \overset{\text{SW13}}&{=} \corr_{\up\wbar{\up\down}}^{\omega_3 \omega_2}
    \\
    \corr_{\wbar{\down}\up\wbar{\up}}^{\omega_1 \omega_2}
    \overset{\text{SW12}}&{=} \corr_{\up\wbar{\down\up}}^{\omega_2 \omega_1}
    \overset{\text{SW23}}{=} \corr_{\up\wbar{\up\down}}^{\omega_2 \omega_3}.
\end{align}
Note that they are not unique. $\corr_{\wbar{\up\down}\up}$, e.g., can also be calculated
from $\corr_{\up\wbar{\up\down}}$ by applying time reversal symmetry. These seven
equations relate seven of the 20 spin components to the $\up\up\down$ and $\up \nbs
\wbar{\up\down}$ components, additionally we have the $\up\up\up$ component. The second
half of the 20 non-vanishing components can be mapped to the these first 10 by using SU(2)
symmetry.
 
    \bibliography{Paper,Main}{}
\end{document}